\input harvmac
\noblackbox
\def\abstract#1{
\vskip .5in\vfil\centerline
{\bf Abstract}\penalty1000
{{\smallskip\ifx\answ\bigans\leftskip 1pc \rightskip 1pc 
\else\leftskip 1pc \rightskip 1pc\fi
\noindent \abstractfont  \baselineskip=12pt
{#1} \smallskip}}
\penalty-1000}
%
\def\hth/#1#2#3#4#5#6#7{{\tt hep-th/#1#2#3#4#5#6#7}}
\def\nup#1({Nucl.\ Phys.\ $\us {B#1}$\ (}
\def\plt#1({Phys.\ Lett.\ $\us  {B#1}$\ (}
\def\cmp#1({Comm.\ Math.\ Phys.\ $\us  {#1}$\ (}
\def\prp#1({Phys.\ Rep.\ $\us  {#1}$\ (}
\def\prl#1({Phys.\ Rev.\ Lett.\ $\us  {#1}$\ (}
\def\prv#1({Phys.\ Rev.\ $\us  {#1}$\ (}
\def\mpl#1({Mod.\ Phys.\ Let.\ $\us  {A#1}$\ (}
\def\atmp#1({Adv.\ Theor.\ Math.\ Phys.\ $\us  {#1}$\ (}
\def\ijmp#1({Int.\ J.\ Mod.\ Phys.\ $\us{A#1}$\ (}
\def\jhep#1({JHEP\ $\us {#1}$\ (}

\def\subsubsec#1{\vskip 0.2cm \goodbreak \noindent {\it #1}\br 
\vskip -12pt}

\def\bb#1{{\bar{#1}}}
\def\bx#1{{\bf #1}}
\def\cx#1{{\cal #1}}
\def\tx#1{{\tilde{#1}}}
\def\hx#1{{\hat{#1}}}
\def\rmx#1{{\rm #1}}
\def\us#1{\underline{#1}}
\def\fc#1#2{{#1\over #2}}
\def\frac#1#2{{#1\over #2}}

\def\br{\hfill\break}
\def\noi{\noindent}

\def\al{\alpha}\def\be{\beta}\def\ga{\gamma}\def\eps{\epsilon}\def\om{\omega}
\def\p{\partial}
\def\IP{{\bf P}}
\def\CY{Calabi--Yau }

\def\la{\lambda}\def\La{\Lambda}\def\Si{\Sigma}

\def\ffc(#1,#2){\ifnum #2=1 {\scriptstyle #1} 
\else {{{1}\over{#2}}{(\scriptstyle #1})} \fi}
\def\eps{\epsilon}
\def\CC{\bx C}\def\ZZ{ {\bx Z}}
\def\kb{\cx K_{\IP^2}}\def\kbff{\cx O(-4)_{\IP^2}}
\def\cf{\cx O(-1)^{\oplus 2}_{\IP^1}}
\def\TT{{\bx T}}
\def\vx#1{ {\vec{#1}} }

\def\fcs(#1,#2){ {\scriptstyle {#1 \over #2}}}
\def\cMb{{\overline \cx M}}
\def\rrr{\ }

\lref\KM{M.~Kontsevich and Y.~Manin,
``Gromov-Witten classes, quantum cohomology, and enumerative geometry,''
Commun.\ Math.\ Phys.\  {\bf 164}, 525 (1994)
[arXiv:hep-th/9402147].}
\lref\AGNT{
I.~Antoniadis, E.~Gava, K.~S.~Narain and T.~R.~Taylor,
``Topological amplitudes in string theory,''
Nucl.\ Phys.\ B {\bf 413}, 162 (1994)
[arXiv:hep-th/9307158].}
\lref\TV{T.~R.~Taylor and C.~Vafa,
``RR flux on Calabi-Yau and partial supersymmetry breaking,''
Phys.\ Lett.\ B {\bf 474}, 130 (2000)
[arXiv:hep-th/9912152].}
\lref\PM{P.~Mayr,
``On supersymmetry breaking in string theory and its realization in brane  worlds,''
Nucl.\ Phys.\ B {\bf 593}, 99 (2001)
[arXiv:hep-th/0003198].}
\lref\guk{S.~Gukov, C.~Vafa and E.~Witten,
``CFT's from Calabi-Yau four-folds,''
Nucl.\ Phys.\ B {\bf 584}, 69 (2000)
[Erratum-ibid.\ B {\bf 608}, 477 (2000)]
[arXiv:hep-th/9906070];\br
S.~Gukov,
``Solitons, superpotentials and calibrations,''
Nucl.\ Phys.\ B {\bf 574}, 169 (2000)
[arXiv:hep-th/9911011].
}
\lref\AVi{M.~Aganagic and C.~Vafa,
``Mirror symmetry, D-branes and counting holomorphic discs,''
arXiv:hep-th/0012041.}
\lref\AVii{
M.~Aganagic, A.~Klemm and C.~Vafa,
``Disk instantons, mirror symmetry and the duality web,''
arXiv:hep-th/0105045.}
\lref\AViii{
M.~Aganagic and C.~Vafa,
``Mirror symmetry and a G(2) flop,''
arXiv:hep-th/0105225.}
\lref\AViv{
M.~Aganagic and C.~Vafa,
``G(2) manifolds, mirror symmetry and geometric engineering,''
arXiv:hep-th/0110171.}
\lref\AVv{
B.~Acharya, M.~Aganagic, K.~Hori and C.~Vafa,
``Orientifolds, mirror symmetry and superpotentials,''
arXiv:hep-th/0202208.}
\lref\quintic{
I.~Brunner, M.~R.~Douglas, A.~E.~Lawrence and C.~Romelsberger,
``D-branes on the quintic,''
JHEP {\bf 0008}, 015 (2000)
[arXiv:hep-th/9906200].}
\lref\app{{\tt http://wwwth.cern.ch/data/app.ps}}
\lref\YZ{
T.~M.~Chiang, A.~Klemm, S.~T.~Yau and E.~Zaslow,
``Local mirror symmetry: Calculations and interpretations,''
Adv.\ Theor.\ Math.\ Phys.\  {\bf 3}, 495 (1999) [arXiv:hep-th/9903053];\br
A.~Klemm and E.~Zaslow,
``Local mirror symmetry at higher genus,''
arXiv:hep-th/9906046.}
\lref\yau{
B.~H.~Lian, K.~F.~Liu and S.~T.~Yau,
``Mirror principle. I,''
{\it  In *Yau, S.T. (ed.): Differential geometry inspired by string theory* 405-454}; ``Mirror principle. II,''
{\it ibid. 455-509}.}
\lref\LMV{
J.~M.~Labastida, M.~Marino and C.~Vafa,
``Knots, links and branes at large N,''
JHEP {\bf 0011}, 007 (2000) [arXiv:hep-th/0010102].}
\lref\MV{M.~Marino and C.~Vafa,
``Framed knots at large N,'' arXiv:hep-th/0108064.}
\lref\LM{J.~M.~Labastida and M.~Marino,
Commun.\ Math.\ Phys.\  {\bf 217}, 423 (2001)
[arXiv:hep-th/0004196].}
\lref\GV{
R.~Gopakumar and C.~Vafa,
``M-theory and topological strings. I,'' arXiv:hep-th/9809187;
``M-theory and topological strings. II,'' arXiv:hep-th/9812127.}
\lref\LN{A.~E.~Lawrence and N.~Nekrasov,
``Instanton sums and five-dimensional gauge theories,''
Nucl.\ Phys.\ B {\bf 513}, 239 (1998)[arXiv:hep-th/9706025].}
\lref\kachru{S.~Kachru, S.~Katz, A.~E.~Lawrence and J.~McGreevy,
``Open string instantons and superpotentials,''
Phys.\ Rev.\ D {\bf 62}, 026001 (2000)
[arXiv:hep-th/9912151];
``Mirror symmetry for open strings,''
Phys.\ Rev.\ D {\bf 62}, 126005 (2000)
[arXiv:hep-th/0006047].}
\lref\bcov{M.~Bershadsky, S.~Cecotti, H.~Ooguri and C.~Vafa,
``Kodaira-Spencer theory of gravity and exact results for quantum string amplitudes,''
Commun.\ Math.\ Phys.\  {\bf 165}, 311 (1994)
[arXiv:hep-th/9309140].}
\lref\Wcs{E.~Witten,
``Chern-Simons gauge theory as a string theory,''
arXiv:hep-th/9207094.}
\lref\Wlsm{E.~Witten,
``Phases of N = 2 theories in two dimensions,''
Nucl.\ Phys.\ B {\bf 403}, 159 (1993)
[arXiv:hep-th/9301042].}
\lref\Kont{M.~Kontsevich,
``Enumeration of rational curves via Torus actions,''
arXiv:hep-th/9405035.}
\lref\GZ{T.~Graber and E.~Zaslow,
``Open string Gromov-Witten invariants: Calculations and a mirror  'theorem',''
arXiv:hep-th/0109075.}
\lref\KL{S.~Katz and C.~C.~Liu,
``Enumerative Geometry of Stable Maps with Lagrangian Boundary Conditions and Multiple Covers of the Disc,''
arXiv:math.ag/0103074.}
\lref\LS{J.~Li and Y.~S.~Song,
``Open string instantons and relative stable morphisms,''
arXiv:hep-th/0103100.}
\lref\OV{H.~Ooguri and C.~Vafa,
``Knot invariants and topological strings,''
Nucl.\ Phys.\ B {\bf 577}, 419 (2000)
[arXiv:hep-th/9912123].}
\lref\vafaln{C.~Vafa,
``Superstrings and topological strings at large N,''
J.\ Math.\ Phys.\  {\bf 42}, 2798 (2001)
[arXiv:hep-th/0008142].}
\lref\ffs{P.~Mayr,
``Mirror symmetry, N = 1 superpotentials and tensionless strings on  Calabi-Yau four-folds,''
Nucl.\ Phys.\ B {\bf 494}, 489 (1997)
[arXiv:hep-th/9610162].}
\lref\osi{P.~Mayr,
``N = 1 mirror symmetry and open/closed string duality,''
arXiv:hep-th/0108229.}
\lref\gmp{B.~R.~Greene, D.~R.~Morrison and M.~R.~Plesser,
``Mirror manifolds in higher dimension,''
Commun.\ Math.\ Phys.\  {\bf 173}, 559 (1995)
[arXiv:hep-th/9402119].}
\lref\amv{A.~Klemm, P.~Mayr and C.~Vafa,
``BPS states of exceptional non-critical strings,''
arXiv:hep-th/9607139.}
\lref\osii{W.~Lerche and P.~Mayr,
``On N = 1 mirror symmetry for open type II strings,''
arXiv:hep-th/0111113.}
\lref\Cgms{C.~Vafa,
``Extending mirror conjecture to Calabi-Yau with bundles,''
arXiv:hep-th/9804131.}
\lref\IqP{
A.~Iqbal and A.~K.~Kashani-Poor,
``Discrete symmetries of the superpotential and calculation of disk  invariants,''
arXiv:hep-th/0109214.}
\lref\JB{
J.~D.~Blum,
``Calculation of nonperturbative terms in open string models,''
arXiv:hep-th/0112039.}
\lref\GJS{
S.~Govindarajan, T.~Jayaraman and T.~Sarkar,
``Disc instantons in linear sigma models,''
arXiv:hep-th/0108234.}
\lref\GP{T. Graber and R. Pandharipande,
``Localization of virtual classes,''
arXiv:alg-geom/9708001.}
\lref\Fab{
C. Faber, 
``Algorithms for computing intersection numbers on moduli spaces of curves, with an application to the class of the locus of Jacobians'',
arXiv:alg-geom/9706006 and MapleV code.}
\lref\Wtsm{E.~Witten,
Commun.\ Math.\ Phys.\  {\bf 118}, 411 (1988);
``Mirror manifolds and topological field theory,''
arXiv:hep-th/9112056.}
\lref\GMM{%
E.~Guadagnini, M.~Martellini and M.~Mintchev,
``Wilson Lines In Chern-Simons Theory And Link Invariants,''
Nucl.\ Phys.\ B {\bf 330}, 575 (1990).}
\lref\MO{P.~Ramadevi and T.~Sarkar,
``On link invariants and topological string amplitudes,''
Nucl.\ Phys.\ B {\bf 600}, 487 (2001)
[arXiv:hep-th/0009188].}

\vskip-2cm
\Title{\vbox{
\rightline{\vbox{\baselineskip12pt\hbox{CERN-TH/2002-062}
                            \hbox{hep-th/0203237}}}\vskip1.5cm}}
{Summing up Open String Instantons}\vskip -1cm
\centerline{\titlefont and N=1 String Amplitudes}
\abstractfont 
\vskip 0.8cm
\centerline{P. Mayr}
\vskip 0.8cm
\centerline{CERN Theory Division} 
\centerline{CH-1211 Geneva 23}
\centerline{Switzerland}
\vskip 0.3cm
\abstract{%
We compute the instanton expansions of the holomorphic couplings 
in the effective action of certain $\cx N=1$ supersymmetric 
four-dimensional open string vacua.
These include the superpotential $W(\phi)$,
the gauge kinetic function $f(\phi)$ and a series of 
other holomorphic couplings which are known to be
related to amplitudes of topological open strings at higher
world-sheet topologies.
The results are in full agreement with the interpretation of the 
holomorphic couplings as counting functions of BPS domain walls.
Similar techniques are used to compute genus one partition function for
the closed topological string on Calabi--Yau 4-fold which 
gives rise to a theory with the same number of supercharges in two dimensions.
}
\vskip1cm
\Date{\vbox{\hbox{ {March 2002}}
}}
\goodbreak

\parskip=4pt plus 15pt minus 1pt
\baselineskip=14pt 
\leftskip=8pt \rightskip=10pt
\newsec{Introduction}
The low energy effective action of many $\cx N=1$ supersymmetric string vacua
is described by a standard four-dimensional supergravity action defined 
by the two functions
$$
f(\phi),\qquad 
G=K(\phi,\bb \phi)+\ln\, W(\phi)+\ln\, \bb W(\bb \phi),
$$
where $\phi$ denotes a chiral $\cx N=1$ multiplet. The K\"ahler potential
$K$ is a real function, whereas the superpotential $W(\phi)$ and 
the field dependent gauge coupling $f(\phi)$ are holomorphic 
in the chiral multiplets.
The computation and interpretation of instanton corrections to 
the physical amplitudes in these phenomenologically relevant 
string vacua has experienced a remarkable development
over the last few years. 
The non-perturbative $\cx N=1$ superpotential 
from genus zero world-sheet instantons
can be determined by now for quite a few
classes of $\cx N=1$ supersymmetric backgrounds by 
generalizations of the idea of mirror symmetry. 
These include type II closed string backgrounds with
fluxes \TV\PM\guk\vafaln\ and type II backgrounds with open string 
sectors from background D-branes. 

In this paper we study the open string case with an focus on the
other holomorphic coupling, the gauge kinetic function $f(\phi)$,
as well as some natural generalizations.
The open string vacua that we consider are type 
II compactifications on Calabi--Yau manifolds with additional
background D-branes. In this case, 
a powerful framework for couplings other then the superpotential exists.
In fact it has been known for quite some time \bcov\AGNT\OV\vafaln\ 
that an infinite number of physical $\cx N=1$ amplitudes in these vacua
are computed by the topological version \Wcs\ of open type II strings.
These are the holomorphic F-terms:
\eqn\topc{
\int d^2\theta h\, N\,\cx  F_{g,h}(\phi)\ \cx W^{2g} S^{h-1}.
}
Here $g$ and $h$ denote the genus and number of boundaries of 
the string world-sheet, respectively,
$\cx W$ is the superfield for the graviphoton 
field strength and $S=tr W_\al W_\al$ is the chiral
superfield for the gauge field on $N$ coinciding  D-branes.
In particular the superpotential $W$ and the $f$-function
are related to the genus zero
partition functions $\cx F_{g=0,h=1}$ and $\cx F_{g=0,h=2}$
with one and two boundaries, respectively. Other world-sheet
topologies may contribute to $W$ and $f$ in the
presence of vev's for the graviphoton field strength
and the gaugino bilinear.

The all genus partition function is related in a beautiful way to a
1-loop Schwinger integral of a dual M-theory compactification 
\GV\OV.
In this context the partition function is the 
weighted counting function of M-theory BPS states.
As the number of BPS states is clearly integer, eq.\topc\ leads to 
the intriguing prediction that the coefficients of 
the instanton expansion of the holomorphic terms in the  
$\cx N=1$ four-dimensional string effective action are 
essentially integral. The unraveling of this structure 
involves a careful treatment of multiple wrappings and their
bound states, studied in \OV\LMV\MV.

Despite the remarkable progress in open string mirror symmetry, 
there are still many open problems. 
The computation of the superpotential  $\cx F_{0,1}$ 
involves a generalization of mirror symmetry to open strings \Cgms\AVi\OV,
which has been a subject of intense studies over the last two years
\kachru\phantom{\AVii\LS\KL\AViii\osi\GJS\GZ\AViv\osii\IqP\JB}
\hskip-170pt-\AVv.  
For other topologies, an explicit open string recursion 
relation in $g$ and $h$ for the partition functions $\cx F_{g,h}$
along the lines of \bcov\ would be desirable\foot{An explicit
expression for the holomorphic anomaly is known for $g=0,h=2$ \bcov.}. 

There is another way to compute the partition functions
$\cx F_{g,h}$, proposed originally by Kontsevich \Kont\ and 
generalized recently to world-sheets with boundaries in \LS\KL\GZ.
The purpose of this note is to explore further the 
structure of the holomorphic $\cx N=1$ string amplitudes 
by an explicit computation of topological partition functions 
$\cx F_{g,h}$ in this framework. 
One of the basic hopes is that the  knowledge
of the instanton expansions, apart from being 
an important physical quantity, will serve 
as a starting point to develop more general
principles that provide a closed form 
of the amplitudes and apply more globally
in the space of perturbative moduli. 

The organization of this note is as follows. In sect. 2 we review the
relation of the physical couplings in the string effective action to 
the partition functions of the topologial string and recall
the interpretation of the latter 
as weighted counting functions of BPS states in 
type IIA/M-theory. In sect. 3 we describe
the D-brane geometry in terms of a linear sigma model (LSM).
We discuss an ambiguity in the definition of the non-perturbative
D-brane geometry which amounts to a choice of a $U(1)$ direction
in the $U(1)^2$ global symmetry of the LSM. 
An geometric interpretation is given that is directly
related to the 
framing in Chern-Simons theory and predicts a specific dependence
of the A-model computation on the integral parameter $\nu$ 
that labels the $U(1)$ direction. 
This is the A-model version of the ambiguity discovered
in the mirror B-model in \AVii\ and studied in 
the context of framings in \MV.
In sect. 4 we describe the localization computation in the A-model for the
two basic one-moduli cases and compute some partition functions $\cx F_{g,h}$
for various phases of D-branes. In particular we compute the 
superpotential $W(\phi)$ and the gauge kinetic functions $f(\phi)$ 
of the string effective $\cx N=1$ action,
as well as some higher genus generalizations thereof.
Re-summing the fractional coefficients of the instanton 
expansions $\cx F_{g,h}$
as predicted by M-theory, leads to an integral expansion
in an extremely non-trivial way, giving a further verification
on the ideas of \OV, as well
as the techniques proposed in \LS\KL\GZ. In sect. 5 we discuss
closed string vacua with the same amount of supersymmetry, 
obtained from a \CY 4-fold compactification. We describe
the localization computation for the 4-fold, which is slightly
different due to the different ghost number of the vacuum, 
and compute the genus zero and genus one partition
functions for the basic one-modulus case. The genus zero result
is in agreement with a computation by local mirror symmetry for
the 4-fold. Some results on the open string amplitudes 
are collected in the appendix, 
an expanded version of which can be found at \app.

\newsec{Open topological strings and counting of BPS domain walls}
In this section we review a few facts and describe the setup 
used in the following sections.
We will mainly consider four-dimensional string vacua with 
$\cx N=1$ supersymmetry obtained from type IIA compactification
on \CY 3-folds $X$ with background D-branes 
wrapping a 3-cycle $L$ in $X$ and filling space-time. 
If $L$ is a special Lagrangian (sL) 3-cycle, the effective 
four-dimensional theory on the brane preserves perturbatively 
$\cx N=1$ supersymmetry. Instanton effects may
generate a non-zero superpotential \Wcs\bcov\quintic\OV\kachru.

To avoid questions of global flux conservation,
we consider non-compact models which describe the local 
neighborhood of a D-brane in a compact manifold $X$. 
The non-compact manifolds will be defined by a linear sigma
model \Wlsm, which allows also for a simple description 
of a class of sL 3-cycles found in \AVi.

The type II D-brane configuration alluded to above is compatible with 
a topological twist of the world-sheet theory of the type IIA string, the 
so-called A-model \Wtsm. The correlation functions of the 
topological A-model on $X$ receive contributions
from holomorphic maps from the world-sheet $\Sigma$ to $X$,
with the boundary $\p \Sigma$ mapped to $L$ \Wcs. The topological 
partition functions $\cx F_{g,h}$ for 
world-sheets with genus $g$ and $h$ boundaries, are related to 
physical amplitudes in the four-dimensional type IIA 
theory by \topc \bcov\AGNT\vafaln\OV.

The partition functions $\cx F_{g,h}$ of the A-model may be
computed by ``counting'' the holomorphic maps from $\Sigma$ to  $X$ 
\Wtsm\Wcs.
A single world-sheet instanton that maps $\Sigma$ to a 
holomorphic curve $C$ in $X$ contributes a term $\exp(-Vol(C))$ 
times a phase factor. More precisely, there are often families
of maps and the ``number'' of maps is replaced by the virtual 
Euler number of the appropriately compactified moduli space.
This has been made precise for world-sheets without boundaries 
in \Kont\GP: 
\eqn\defkd{
\cx F_{g,0} = \sum_{g,\be} F_{g,\be}\,  q(\be),\qquad
F_{g,\be}= \int_{\cMb^{vir}_{g,0}(\be,X)} c_{top}(U_\be).
}
Here $\cMb^{vir}_{g,k}(\be,X)$ is the virtual 
moduli cycle for stable maps 
from genus $g$ curves with $k$ marked points into $X$,
$\be\in H_2(X,\ZZ)$ the class of the image and $q(\be)$
its complexified K\"ahler volume. Moreover $c_{top}(U_\be)$ is the
Euler class of the relevant obstruction bundle $U_\be$.
The fractional coefficients $F_{g,\be}$ are 
the Gromov--Witten (GW) invariants. 
A, somewhat preliminary, generalization of these definitions
to world-sheets with boundaries 
has been developed recently in the papers \LS\KL\GZ.
It will be used in the following to determine the partition 
functions $\cx F_{g,h}$. The highly non-trivial consistency 
of the results may serve as a further verification 
of these generalizations.

A strong consistency condition arises
from the interpretation of $\cx F_{g,h}$ as a
type IIA/M-theory  1-loop amplitude 
that computes the effective couplings \topc\ in a constant 
self-dual graviphoton background in two dimensions \GV\OV.
The amplitude receives contributions only from short BPS 
multiplets and predicts a re-summation of $\cx F_{g,h}$ 
into a weighted counting function of integral degeneracies of
BPS super-multiplets in a given representation\foot{For 
earlier work along these lines, see \LN.}.
For a single stack of $N$ D-branes on a sL 3-cycle $L$ 
with $h_1(L)=1$, the instanton expansion of 
the partition function (for a fixed number of boundaries) is of the form 
\eqn\gpf{
\cx F_{h}(t_i,r',V) =  \sum_{g;w_\al} g_s^{2g-2+h} \fc{1}{h!}
F_{g;w_\al}(q_i) e^{-2\pi r' w}\, \prod_{\al=1}^h \rmx{Tr}\, V^{w_\al},
}
where $g_s$ is the string coupling constant. Moreover
the $q_i=exp(2\pi i t_i)$ are the exponentials of the 
closed string moduli, $r'$ is the K\"ahler volume of the
primitive disc with boundary on $L$ and $V$ is a diagonal 
$N\times N$ matrix with entries $V_{aa}=e^{i\phi_a}$,
where $\phi_a$ is the holonomy in the $a$-th $U(1)$ factor 
along the non-trivial cycle in $L$. Moreover the integer $w_\al$
is the winding number of the $\al$-th boundary along the $S^1\subset L$
and $w=\sum_\al w_\al$. 

The re-summation of the partition function 
$\cx F_{g,h}$ in terms of multiplicities of BPS
states has the form \LMV\MV:
\eqn\mcf{
\sum_{g=0}^{\infty}g_s^{2g-2+h}F_{g,\be}=
\fc{(-1)^h }{\prod_\al w_\al} \, \sum_{\be'=\be/d} n_{g,\be'}\ d^{h-1}
\big(2 \sin \fc{dg_s}{2}\big)^{2g-2}\prod_\al
(2\sin \fc{w_\al g_s}{2})\, q(\be),
}
where $q(\be)$ is again the exponentiated complex K\"ahler volume
of the class $\be\in H_2(X,L)$. The
class $\be$ is specified by the winding numbers $w_\al$ of the boundaries
of disc components 
and $h^{1,1}(X)$ degrees $n_i$ for a basis of 
holomorphic 2-cycles $C_i$ for $H_2(X,\ZZ)$.
The coefficients $n_{g,\be}$ are integral linear combinations\foot{See
\LMV\ for more details.} of
the number of $\cx N=1$ supersymmetric BPS multiplets of a 
given bulk charge determined by the class $\be$.
The terms with $d>1$ in the above formula are the 
contributions from multiply wrapped branes, 
which have an interpretation as the 
momentum states along the
extra circle in the M-theory compactification 
\LN\GV\foot{The class $\be'=\be/d\in H_2(X,L;\ZZ)$ exists, 
if all the degrees $n_i$ and windings $w_\al$ specifying the class 
$\be$ are are divisible by $d$.}.

The partition function $\cx F_{h}$ as written in \gpf\ is the 
specialization $\rmx{Im}\, t'_a=r'\, \forall \, a$
of a holomorphic section $\cx F_{h}(t'_i,t'_a)$,
where $t_i$ and $t'_a$ are the bottom components 
of the moduli superfields from the closed and 
open string sector, respectively. 
It comprises amplitudes
from various types of world-sheets that involve $N'\leq N$ branes.
To extract these different sectors and to make the relation to the
results obtained from the following A-model computations explicit\foot{%
This has been also studied by M. Mari\~no and E. Zaslow 
\ref\MZ{Private communication.}.},
it is useful to restore the dependence on the 
complex open string moduli $t'_a$. The moduli $t_i$ and $t'_a$ 
are defined classically by the action of the primitive world-sheet 
instantons \Wtsm\Cgms:
\eqn\defmodi{
t_i = \int_{C_i} b + i J,\, i=1,...,h^{1,1}(X),\qquad 
t'_a = \int_{\ga_a} A + \int_{D_a} iJ, \,  a=1,...,N.
}
Here $J$ is the K\"ahlerform, $b$ the anti-symmetric 2-form and $A$ the
gauge field on the D-brane. Moreover the $D_a$ are $N$ 
discs ending at the $N$ D-branes on the 1-cycles $\ga_a$.
The imaginary parts $\rmx{Im}\, t'_a$ parametrize the 
$N$ independent positions of the D-branes that determine also the size
of the discs $D_a$. The dependence of the sections
$\cx F_{g,h}$ on the complex moduli is determined by holomorphicity:
\eqn\gpfii{
\cx F_{g,h}(t_i,t'_a)= \sum_{g;w_\al} g_s^{2g-2+h} \fc{1}{h!}
F_{g;w_a}(q_i) \sum_{\vx a} \prod_{\al=1}^h (v_{a_\al})^{w_\al}.
}
In the above $v_a = exp(2\pi i t'_a)$ and the last sum
is over $N\cdot h$ choices $a_\al \in \{1,...,N\}$.
The partition functions $\cx F_{g,h}$ for $N'\leq h$ branes 
can be obtained by moving $N-N'$ of the branes to infinity.
This is the limit $v_a=0,\, a=N'+1,...,N$. 
From the above it is clear that the independent GW invariants 
are related to  world-sheets with $h$ distinguished 
boundaries landing on $N=h$ different branes.
This is the siutation which we will consider in the A-model computation.
The multi-cover structure will be extracted from the specialized
form \mcf\ for a stack of  $h=N$ D-branes.
Partition functions $\cx F_{g,h}$ and their 
invariants for less then $h$ branes are obtained from those
for $N=h$ branes by the appropriate identifications amongst 
the open string moduli.

\newsec{A-type D-branes and torus actions}
In the following we discuss the definition of the A-model D-brane geometry 
and its moduli space in terms of a gauged LSM. An important detail
is the existence of an integral parameter $\nu$ that enters the
non-perturbative definition of the string vacuum. This is 
the A-model version of the framing ambiguity in the 
Chern-Simons theory \MV, and the dependence of the B-model on
a certain boundary condition \AVii. 
Here we propose a simple geometric
interpretation of the ambiguity in the A-model which implies the 
specific dependence of the partition function
on this parameter.

The classical D-brane geometry for the A-model will be defined
as follows. We consider  \CY 3-folds $X$ defined as a direct sum
of concave line bundles\foot{For a discussion and classification of more
general bundles, see \yau.} over $\IP^n$. Much of the following
discussion holds generally for D-branes of the same topology.
In concrete
we will study the two one-moduli cases 
$\kb$, the canonical bundle of $\IP^2$ and the 
bundle $\cf$ that describes the local geometry 
of the blow up of the conifold.
The manifold $X$ may be
defined as a  2d linear sigma model \Wlsm, a (2,2) supersymmetric
$U(1)$ theory with four matter fields $Z_i$ of charges
$(1,1,1,-3)$ for $\kb$ and $(1,1,-1,-1)$ for $\cf$, respectively.
The solution to the D-term vacuum equations in the scalar components $z_i$
$$\eqalign{
\kb:&\ |z_1|^2+|z_2|^2+|z_3|^2-3|z_4|^2-r=0,\cr
\cf:&\ |z_1|^2+|z_2|^2-|z_3|^2-|z_4|^2-r=0.}
$$
describe a smooth \CY 3-fold $X$ for large positive 
values of the real FI parameter $r$. The single 
complex K\"ahler modulus of $X$ is $t=b+ir$ with
$b$ the value of the $B$-field on the fundamental 
sphere in $X$.

In addition the type II string vacuum includes D-branes
on sL 3-cycles $L$ in $X$. As in \AVi\ we consider 3-cycles of
topology $S^1\times \CC \simeq S^1\times \tx S^1
\times \bx R_{\geq 0}$, defined by the equations
\eqn\defdb{
|z_3|^2-|z_2|^2=c^1,\ \  
|z_3|^2-|z_4|^2=c^2,\ \  
Arg(z_1z_2z_3z_4)=0.
}
Here $c^a$ are two complex constants that are 
chosen such that one of the $S^1$ factors (representing
$\tx S^1 \subset \CC$) shrinks over some edge $z_i=z_j=0$. 
In the relevant patch of $X$,  let $z_a,z_b$ denote the two gauge invariant
coordinates that vanish at the origin of
$\CC\subset L$ and 
$z_D$ the third gauge invariant coordinate on the 3-fold $X$.
The primitive disc in this geometry, that enters the classical definition of
the open string modulus \defmodi\ is defined as 
\eqn\defdisc{
D:\ z_a=z_b=0,\ |z_D|^2\leq c.
}
with $c$ some complex constant. 

The sL cycle $L$ is a cone over $T^2$, where the 1-cycle $\tx S^1\in\CC$
shrinks at the tip of the cone. 
The complex torus base can be identified with an
$U(1)^2$ subgroup of the global symmetry group $U(1)^3$ of the gauged LSM.
The latter acts by the rotations $(z_a,z_b,z_D)\to$
$(z_a\, e^{i\la_a\varphi},z_b\, e^{i\la_b\varphi},z_D\, e^{i\la_D\varphi})$
on the inhomogeneous coordinates. The sL condition \defdb\ selects
the $U(1)^2$ subgroup,
\eqn\cons{
\la_a+\la_b+\la_D = 0,}
which is the same as the condition of anomaly freedom for a
gauged $U(1)$ symmetry. 

There is an inherent ambiguity in the definition of the
non-perturbative D-brane geometry which amounts to the choice
of a $U(1)$ subgroup of the global symmetry group. We choose
coordinates $z_a$ and $\arg(z_D)$ on $L\subset X$ and denote by
$(\ga_0,\ga_1)$ a basis of $H_1(T^2)$ parametrized by the
phases $(\theta_a,\theta_D)$. Although $\ga_0$ is homologically
trivial, it will become relevant when defining
a pair of non-trivial loops in $L$. E.g. the geometry of 
a single world-sheet instanton ending on the D-brane wrapped on $L$
is defined by two maps $f:\, (\Si,\p \Si) \to (\Si,L)$ and 
$g:\, W \to L$, where $\Si$ is the world-sheet and $W$ the
part of the D-brane world-volume mapping into $X$. In the definition
of the pair of maps $f$ and $g$ one has to  specify how often the 
image of a fixed generator in $H_1(W)$, defined by the non-trivial 
Wilson line, wraps around the image of $\p \Si$. 
The image of the boundary $\p \Si$ of a minimal 
volume world-sheet is in the class $[\ga_1]$ and the image
of $H_1(W)$ is in the class
\eqn\dbbound{
[\ga(\nu)] = 1\cdot [\ga_1] + \nu \cdot [\ga_0],\qquad \nu \in \ZZ,
}
where the coefficient $\nu$ of the trivial cycle $\ga_0$ is 
well-defined in the presence of the world-sheet instanton.
The shifts of the origin of the 1-cycle $\ga(\nu)\simeq S^1$ define a
$U(1)$ subgroup of the global $U(1)^2$ symmetry with weights 
\eqn\defnu{\la_a= \nu \la_D,\qquad \la_b=-(\nu+1)\la_D.}
The above geometric picture is closely related to the definition
of a framing in the context of Chern-Simons theory. The latter
is a choice of UV regularization for the product of 
Wilson line operators on the same knot $k$ which preserves general
covariance. It is specified by a choice of  a normal vector to $k$
and the inequivalent choices are labeled by an integer $\nu$ 
that counts how often the first knot wraps the second one in
transverse space. In the above, the transverse space (within $W$) 
to the non-trivial
homology cycle $\ga_1$ is $\CC$ and thus the integer $\nu$ indeed defines
a framing for a Wilson loop on it.

The following A-model computations of the partition functions 
$\cx F_{g,h}$ are based on counting the ``number'' of discs
with boundary on $L$. More precisely this number is defined as
an Euler number of the appropriately defined moduli space
of discs. The phases of the open string moduli \defmodi\ are
the Wilson lines along $H_1(W)$ and the definition of 
the instanton corrected moduli space depends on the 
invariant $\nu$ for the maps $f$ and $g$ described above\foot{A similar
comment applies to several loops arising from world-sheet instantons
with more boundaries. One may think of the topological
instantons as the BPS limit of non-minimal world-sheet wrappings 
with boundary in the class $\ga$.}.
On the other hand the computation is technically performed by 
localizing to the fixed points of the torus action on the moduli
space induced by an $U(1)\subset U(1)^2$ action on $X$. The
deformations for a given choice of non-perturbative D-brane 
geometry are consistent with the specific torus action \defnu.
Thus the A-model computation with torus action \defnu\ computes
the partition function of the D-brane geometry defined by 
$\nu$, which in turn is directly identified with a framing in 
CS theory. The existence of an integral ambiguity in the
A-model computation had been already verified in \KL,
guided by the B-model result of \AVii.

The previous geometric interpretation of the parameter $\nu$ in the A-model
is similar to the one discussed in \AVii\ for a
dual type IIB compactification on a web of $(p,q)$ 5-branes.
It is likely that the geometry of the A-model 
can be used to derive also the functional form of the $\nu$ dependence
of the special coordinate $t'$, obtained in \AVii\ 
in the mirror B-model.

\newsec{Localization in the open string A-model and graph sums}
The method for the computation of $\cx F_{g}$ proposed 
in \Kont\ uses a group $\bx T$ of torus actions on the target space $X$
to localize the integrals \defkd\ to the fixed points of $\bx T$.
The toric manifolds arising from the linear sigma model
have, by definition, a sufficient number of torus actions.
Specifically, the torus $\bx T\simeq (\CC^*)^{4}$ 
acts on the bundle $X\to\IP^n$ by phase rotations of the 
homogeneous coordinates $z_i$.

Let $M$ denote some moduli space of stable maps 
involved in the integral \defkd, $\phi$ the relevant top form
on it  and $M^\TT$
the fixed locus of the induced action of $\TT$ on $M$.
An application of the Atiyah-Bott fixed point theorem
yields the formula (derived for $g>0$ and $h=0$ in \GP)
\eqn\abo{
K=\int_M \phi = \sum_{\cx M^\Gamma}\int_{\cx M^\Gamma} \fc{i^* \phi}
{e(N^{vir})}.}
It localizes the integral \defkd\ to the components $\cx M^\Gamma$ of the
fixed set $M^\TT$ of the torus action. In the above, 
$i:\cx M^\Gamma\hookrightarrow M$ is the embedding map and 
$e(N^{vir})$ the Euler class of the virtual 
normal bundle  $N_{\cx M^\Gamma/M}$.

The space of maps fixed under the torus action has the 
following structure \Kont\KL\GZ. Let $p_i$ denote the fixed points 
$z_k=0,\, k\neq i$ in the base $\IP^n$ of the bundle $X$. 
Consider a domain curve $\Si$ which 
is a union of irreducible components 
$(\cup_\al C_\al)\cup(\cup_\al D_\al)$, where $C_\al$ is a genus
$g_\al$ Riemann surface with $n_\al$ marked points and $D_\al$ 
are disc components with one marked point in the interior.
The irreducible components attach to each 
other at marked points to form the nodal domain curve $\Si$. An invariant map contracts 
all components of $\Si$ to the fixed points $p_i$ with two exceptions:
$i)$ a genus zero component $C_\al$ with two marked points 
maps to the line 
$l_{ij}:\ z_k=0,\, k\neq i,j$ connecting the fixed points $p_i$
and $p_j$ with the marked points mapped to the fixed points; $ii)$
at generic moduli of $X$, the map $f$ restricted to a disc 
component is non-constant and has as its image the disc
$\tx l_{ia}:\, z_k=0,\, k\neq i,a;\, |z_a|^2\leq 1$. Here 
$z_a$ may be either a coordinate $z_i$ on the base 
$\IP^n$, so $\tx l_{ij}$ lies on the line $l_{ij}$
for some $j$, or a coordinate on the fiber of the bundle $X$ over $\IP^n$.
The circle $\gamma_{a,I}$
at $|z_a|^2=1$ parametrized by the phase of $z_a$ 
carries a label $I$ that specifies 
the $I$-th D-brane on which the boundary of the disc lands.
Note that the $D_\al$ are the
only components of $\Si$ which may map outside the
compact base of the bundle $X$. The above discussion
slightly generalizes the set of fixed loci with respect to \KL\GZ.

The irreducible 
components $\cx M^\Gamma$ of the fixed locus $\cx M^\TT$ may be
characterized by decorated graphs $\Gamma$ for world-sheets without
boundaries \Kont\ and a generalization that includes disc components
has been proposed in \GZ. A graph $\Gamma$ will be defined 
by three sets,  the vertices $V$, connected by edges $e\in E$ and 
in addition a set of legs $l\in L$ originating at the vertices. 
A vertex $v\in V$ represents a contracted component or a 
pole of a $\IP^1$ component and carries 
two labels $i(v)$ and $g(v)$ that specify its image in $X$, namely 
the fixed point $p_i$, and the genus of a contracted component.
An edge $e$  represents a non-constant mapping from a genus
zero component $C_\al$  and carries the labels $i(e),j(e)$ and $\vx n$
that specify the image of the south and north poles in $X$ and
the class $\be_e$ in $H_2(X,\ZZ)$ of the image.
The legs represent the disc components and carry 
labels $i(l)$ and $\vx w$ specifying the image of the 
disc center in $X$ and the  class of the image 
as an element of the relative homology $H_2(X,L)$.
In particular $\vx w$ specifies the D-brane label $I$
and the circle $\gamma_{a,I}$.

According to the above discussion, we consider the following
graphs for the computation of the Gromov--Witten invariant
$F_{g,\be}$.  A graph $\Gamma$ represents a component of the
moduli space of maps from world-sheets 
of genus $g$, with $k$ marked points and $h$ disc
components $D_\al$ with image in the class $\be$ if:
\eqn\graphs{
\vbox{\offinterlineskip\tabskip=0pt\halign{
\strut$#\qquad \hfil~$ & $#$\hfil~\cr
1)\ 1+|E|+\sum_V (g(v)-1)= g;
&4)\ \sum_{e\in E} \be_e +\sum_{l\in L} \be_L = \be;\cr
2)\ e\in E\ \Rightarrow \ i(e)\neq j(e);
&5) \cup_v S(v) = \{1,...,k\}.\cr
3)\ l\in L\ \Rightarrow \ \exists\,  \tx l_{i(l)a};&\cr
}}}
Here $S(v)$ is the set of marked points on the component $C_v$ and
$\be_e$ and $\be_l$ are the classes in $H_2(X,L)$ of the images of the 
components $C_e$ and $C_l$, respectively.

The moduli space $\cx M^\Gamma$ is the quotient
$\prod_{v\in V} \cMb_{g(v),val(v)}/Aut(\Gamma)$ of 
the product of moduli 
spaces of Riemann surfaces of genus $g$ and with $val(v)$
marked points. Here $val(v)$ is the number of marked points of the
component $C_v$ associated to the vertex $v$, including the points 
of intersection with the edge and leg components.
As described in the following section, 
the integrand in \abo\ is a formal sum in the classes 
$c_i$ and $\psi_i$ in $H^*(\cMb_{g(v),val(v)},\bx Q)$. 
Here $c_i$ are the Chern classes of the dual of the Hodge bundle
and the $\psi_i$ the first Chern classes of the line bundles 
associated to the marked points $x_i$, with fiber 
the cotangent space at $x_i$. 
The integrals over this sum are computable by Faber's algorithm \Fab. 
The result has to be divided by a
symmetry factor $A_\Gamma$ that takes into
account the quotient structure of $\cx M^\Gamma$. 
It is 
$
A_{\Gamma}=\prod_{l\in L}w^{-1}_l\times {\prod_{e\in E} n_e^{-1}} 
\times a_{\Gamma}
$
where $a_{\Gamma}$ is the order of the automorphism group
of $\Gamma$ as a dressed graph with 
distinguished legs\foot{Note that
$a_\Gamma$ is not equal to the 
symmetry factor for the graph $\Gamma$
with the distinguished legs deleted.} and $n_e=gcd(\vx n(e))$.

\subsec{The integrand on $\cx M^\Gamma$}
It remains to determine the $\TT$ equivariant 
class of the integrand in the integrals of the graph sum \abo.
The obstruction sequence is
$$\eqalign{
0\to Ext^0(\Omega_\Si(E),\cx O_\Si)\to &H^0(\Si,\p \Si; f^*TX,f^*TL) \to \cx T^1\to\cr
Ext^1(\Omega_\Si(E),\cx O_\Si)\to &H^1(\Si,\p \Si; 
f^*TX,f^*TL) \to \cx T^2\to 0,}
$$%
where $E$ is the divisor of marked points on $\Si$.  
The  equivariant class of the integrand $i^*\phi/e(N_{\cx M^\Gamma/M})$ 
is equal to the class $\cx T^2/\cx T^1$ \GP\LS\KL\GZ.
For an explicit computation one may use the normalization of $\Si$ in 
terms of its irreducible components$$
0 \to \cx O_\Si \to \oplus_\al\cx O_{C_\al}\oplus_\al \cx O_{D_\al} \to
\oplus_i TX|_{f(x_i)} \to 0,
$$%
where $x_i$ are the nodes on $\Si$.
The case without boundaries has been considered in \Kont\GP,
and the contribution from the disc components in \LS\KL\GZ.
In particular the leg contribution to the integrand of a 
specific sL 3-cycle $L$ in the conifold \OV\ has been derived in \LS\KL,
and for a phase of a D-brane on $\kb$ in \GZ.
The integrands needed in the following computations are 
obtained by a straightforward though somewhat lengthy 
variation of the arguments in \GP\LS\KL\GZ\ and we refer
to these references for more details on the computation. 
Here we limit ourselves to point out the universal
form of the disc contributions which leads to a 
quick way to determine the open string integrand
for any D-brane phase in any toric \CY 3-fold.

Indeed it is straightforward to see that the addition of disc
components to $\Si$ leads to an essentially universal 
modification of the closed string computation. 
Adding discs to $\Si$ is very similar to adding marked 
points (as discs can not concatenate), up to some 
extra contributions from the characteristic classes related 
to the disc components\foot{This was used in \GZ\ to derive
a mirror identity for $g=0$, $h=1$.}. 
However the latter modifications are 
essentially fixed by the degree zero result $\vx n(e)=0\, \forall\, e\in E$, 
which is universal. This is because in the limit 
of large K\"ahler moduli of $X$, any D-brane 
configuration reduces to that in $\bf C^3$ 
and this is what is described by the partition function of $X$ 
at degree zero.

The contribution to the closed string integrand 
from
the normal bundle is \GP\foot{Below it is
understood that a subscript $i$ refers to an edge label $i(e)$
under the first product and a vertex label $i(v)$ under the
second product. A similar convention will be used in the following
formulae.}
\eqn\intcs{
\eqalign{
\fc{1}{e(N^{vir})}=
&\prod_{e\in E} \big( \fc{(-)^d}{d!^2}(\fc{d}{\la_{i}-\la_{j}})^{2d}
\prod_{k\neq i,j \atop a=0}^{a=d} (\fc{a}{d}\la_{i}+
\fc{d-a}{d}\la_{j}-\la_k)^{-1}\big) \times\cr
&\prod_{v\in V} \big(\prod_{k\neq i}\la_i-\la_k\big)^{val-1} \times 
\cases{(\sum_{F\ni v} w_F^{-1})^{val-3}\ \prod_{F\ni v}w_F^{-1}&$g=0$\cr 
\fc{\prod_{k\neq i} P_g(\la_i-\la_k)}{\prod_{F\ni v} w_F-\psi_F} 
&$g>0$}
}
}
The weights $\la_i$ specify the torus action 
$z_i\to e^{i\la_i \al}z_i$ on the homogeneous coordinates 
of the $\IP^n$ base.
The polynomial $P_g(\la)=\sum_{k=0}^g \la^k c_{g-k}$ is the equivariant 
top Chern class of the dual of the Hodge bundle, twisted by  
$U(1)$ with weight $\la$. 
Moreover a flag $F:(v,e)$ is defined as an 
oriented edge $e$ with origin $i(e)=i(v)$ and 
its weight under the torus action is defined as 
$w_F= (\la_{i(e)}-\la_{j(e)})/n_e$.
A similar expression describes the 
pull-back $i^*\phi$ of the Euler class of the 
obstruction bundle, which depends however on $X$ and will 
thus be stated later. 

Adding disc components to the world-sheet amounts to 
replacing a closed string graph $\Gamma$ by
a related graph $\Gamma'$ with some legs added. The
modified integrand that includes the contributions from the disc
components can 
be conveniently written in two factors. The first one is identical to 
$i^*\phi\, \cdot\, $\intcs,
with $val(v)$ counting the marked points, edges {\it and}
legs at the vertex $v\in V$ of the graph $\Gamma'$ 
and similarly the set $\{F\ni v\}$ of flags  
runs over flags associated with both, edges and legs, attached to $v$.%
\foot{The weight of a disc flag $F:\, (v,l)$  
is defined as $w_F=-\la_D/w(l)$, where $\la_D$ is the weight
of the gauge invariant coordinate $z_D$ on the disc in $X$.}
This is the result of \GZ\ for $\kb$, which holds 
generally for toric 3-folds by the previous universality argument. 
The second factor of the open string integrand 
is the following product of universal disc contributions
\eqn\intL{\eqalign{
&\prod_L \big( \fc{-1}{w!}(\fc{w}{\la_D})^{w}
\big)\ \times \
\big( \prod_{k=1}^{w-1} \la_\bot+\fc{k}{w} \la_D \big)\cr
&=\,(-1)^h\, \prod_L\fc{1}{\la_D} 
\prod_{k=1}^{w-1}1+\fc{w}{k}\fc{\la_{\bot}}{\la_D},
}}
and it describes the contribution from $H^k(D_{\al(l)},f^*TX),
\, k=0,1$. The above expression summarizes the results of \LS\KL\GZ\
in a universal form, the important point being that the
contribution from the cohomology groups that contribute to 
\intL\ depends only on the local disc geometry and is
therefore universal, that is independent of the bundle $V$.

The weight $\la_D$ in \intL\ is the weight of the 
coordinate $z_D$ on the primitive disc in $X$ defined
as in \defdisc. It remains to specify the weight $\la_{\bot}$.
First note that
on dimensional grounds, the integral in \abo\ is of total degree zero 
in the torus weights $\la_i$. The partition function 
$\cx F_{g,h}$ depends therefore only on ratios of
the $\la_i$. The definition 
of the sL cycle $L$ requires the sum of the weights to be zero and 
reduces the free parameter to a single ratio of weights.%
\foot{The closed string partition functions $\cx F_{g,h=0}$
are in fact completely independent of a choice of torus action,
essentially since the closed string
observables are related to the ordinary cohomology 
of $X$.}

As discussed in sect. 3\rrr, the relevant ratio originates in 
the definition of the geometry of the world-sheet instantons,
where we have to specify how often the image of $H_1(W)$ wraps 
around the origin of $\CC\subset L$ when it wraps once around the 
non-trivial $S^1\subset L$. With the coordinates 
$z_a,z_b$ defined as in \defdisc, the geometry with fixed winding $\nu$ 
leads to two possible choices for the $U(1)$ action,
depending on the orientation of the D-brane:
$$
\la_\bot=\la_{z_a}= \nu \la_D,\qquad \la_\bot = \la_{z_b}=\nu' \la_{D},
\qquad \nu \in \ZZ.
$$
The two choices are related by the invariance condition  
as in \defnu. There will be a symmetry of the invariants 
$n_{g,\be}(\nu)$ for the moduli spaces (not the partition functions) 
under the exchange $\nu\to -(\nu+1)$, if 
the exchange of coordinates $z_a\leftrightarrow z_b$ induces
also a symmetry transformation of the closed string background.

\subsec{Case I: The blow up of the conifold $\cf$}
The first one modulus case we consider is the blow up of the conifold.
The degree zero contribution to $\cx F_{g,h}$ has been 
studied in \KL\LS; the same result had been previously obtained
in \OV\ by a Chern-Simons computation for the unknot.
Substantial generalizations of the CS computation to
other world-sheet topologies and 
other knots and links have appeared in 
\LM\LMV\MV\MO. In particular some of the amplitudes computed
below have been already studied explicitely in the Chern-Simons
theory \MV\ for general framing. The partition function 
$\cx F_{0,1}$ has also been obtained in \AVi\MV\ by mirror symmetry.

In the patch $z_1 \neq 0$, the gauge invariant 
coordinates are $z=z_2/z_1$ for the $\IP^1$ and two coordinates  
$a=z_3z_1$ and $b=z_4z_1$ on the $\cx O(-1)$ fibers. 
With the above conventions, the weight of $z$ is $\la_z=\la_2-\la_1$. 
The weights $\la_{a,i},\la_{b,i}$ of the torus action on the fiber 
can be chosen arbitrary at the fixed point $p_1$ and are
related by the projective action on $\IP^1$ to the weights at the 
second fixed point. Specifically $\la_{a,2}=\la_{a,1}+\la_z$, 
and similarly for $b$.

The class of the pull back $i^*\phi$ of the 
Euler class of the obstruction bundle computes to
\eqn\istarcf{
\eqalign{
i^*\phi= 
&\prod_{v\in V} P_g(-\la_{a,i}) P_g(-\la_{b,i}) 
(\la_{a,i}\cdot \la_{b,i})^{val(v)-1}\cr
\times 
&\prod_{e\in E} \prod_{m=1}^{d-1}\big(\la_{a,i}-\fc{m}{d}(\la_i-\la_j)\big)
\big(\la_{b,i}-\fc{m}{d}(\la_i-\la_j)\big).}
}
The open string integrand is the product \intcs $\cdot$ \intL $\cdot$ \istarcf.
The invariance condition for the 
sL cycle \defdb\ reads  $\la_{a,1}+\la_{b,1}+\la_z=0$.
There are two different phases of D-branes,
depending on whether the center of $\CC \subset L$
maps to a point on the compact $\IP^1$ or not. 
In the first case, the primitive disc is $D:\, a=b=0,\, |z|^2\leq c$
and thus $z_D=z, z_{\bot}=a, \nu=\fc{\la_{a,1}}{\la_z}$. In the
second case, $D:\, b=z=0,\, |a|^2\leq c$ and 
$z_D=a, z_{\bot}=z, \nu=\fc{\la_z}{\la_{a,1}}$.
Although the invariants are significantly different in the
two phases\foot{The two phases are related to the two sides of 
a flop transition in a \CY 4-fold by an open/closed string
duality \osi.}, the
the general structure is similar and 
the following discussion will be limited to the second phase.
Results for the first phase can be found in the appendix.

\subsubsec{4.2.1.\rrr The superpotential and higher genus generalizations}
We consider first the superpotential $W=\cx F_{0,1}$ 
on $X$ and its higher genus generalizations described in \topc.
The topological partition functions $\cx F_{g,h=1}$ are obtained 
by evaluating the graph sum \abo\ for graphs with one leg. 
Let $q=exp(2\pi i t)$ denote the exponential of the 
complexified K\"ahler volume of the primitive sphere in 
$X$ and $v_1=exp(2\pi i t')$ the exponential of the single open
string modulus. 
The first terms of the instanton expansion
of the partition functions $\cx F_{g,1}$ for $g<5$ are:
\def\ffc(#1,#2){\ifnum #2=1 {\scriptstyle #1} 
\else {\scriptstyle {{1}\over{#2}}(#1}) \fi}
\eqn\cfpfi{\eqalign{
\cx F_{0,1}=&(v_1+\ffc(1-2 \nu,4) v_1^2+\ffc((-1+3 \nu) (-2+3 \nu),18) v_1^3+\ffc(-(-1+2 \nu) (-1+4 \nu) (-3+4 \nu),48) v_1^4)+\cr
&q_1 (-v_1+\ffc(\nu,1) v_1^2+\ffc(-(-1+3 \nu) \nu,2) v_1^3+
\ffc((-1+2 \nu) (-1+4 \nu) \nu,3) v_1^4)+\cr
&q_1^2 (\ffc(-2 \nu-1,4) v_1^2+\ffc(\nu (3 \nu+1),2) v_1^3+\ffc(-(-1+4 \nu)
 \nu (4 \nu+1),4) v_1^4)+...\cr
\cx F_{1,1}=&
(\fcs(1,24) v_1+\ffc((\nu^2-\nu-1) (-1+2 \nu),24) v_1^2+\ffc(-(2 \nu^2-2 \nu-1) 
(-1+3 \nu) (-2+3 \nu),48) v_1^3
)+\cr
&q_1 (-\fcs(1,24) v_1+\ffc(-(-1+2 \nu^2) \nu,12) v_1^2+\ffc((-5-6 \nu+
18 \nu^2) (-1+3 \nu) \nu,48) v_1^3
)+\cr
&q_1^2 (\ffc((2 \nu+1) (\nu^2+\nu-1),24) v_1^2+\ffc(-\nu (3 \nu
+1) (18 \nu^2+6 \nu-5),48) v_1^3)
\cr
\cx F_{2,1}=&
(\fcs(7,5760) v_1+\ffc(-(3 \nu^4+11 \nu-6 \nu^3-8 \nu^2+7) (-1+2 \nu),1440) v_1
^2
)+\cr
&q_1 (-\fcs(7,5760) v_1+\ffc((-10 \nu^2+3+6 \nu^4) \nu,720) v_1^2
+\ffc(-(108 \nu-276 \nu^2+53-288 \nu^3+432 \nu^4) (-1+3 \nu) \nu,3840) v_1^3
)+\cr
&q_1^2 (\ffc(-(2 \nu+1) (3 \nu^4+6 \nu^3-8 \nu^2-11 \nu+7),1440) v_1^2
+\ffc(\nu (3 \nu+1) (432 \nu^4+288 \nu^3-276 \nu^2-108 \nu+53),3840) v_1^3
)+...\cr
\cx F_{3,1}=&
(\fcs(31,967680) v_1+\ffc((-57 \nu-9 \nu^5-31+45 \nu^3+3 \nu^6-15 \nu^4+33 \nu^2) (-\
1+2 \nu),60480) v_1^2
)+\cr
&q_1 (-\fcs(31,967680) v_1+\ffc(-(21 \nu^2-5-21 \nu^4+6 \nu^6) \nu,30240) v_1^2
)+\cr
&q_1^2 (\ffc((2 \nu+1) (3 \nu
^6+9 \nu^5-15 \nu^4-45 \nu^3+33 \nu^2+57 \nu-31),60480) v_1^2
)+...\cr
\cx F_{4,1}=&
(\fcs(127,154828800) v_1+\ffc(-(741 \nu+190 \nu^5+381-656 \nu^3-20 \nu^7-40 \nu^6+
158 \nu^4+5 \nu^8-378 \nu^2) (-1+2 \nu),7257600) v_1^2)+\cr
&q_1 (-\fcs(127,154828800) 
v_1+\ffc((126 \nu^4-100 \nu^2+21-60 \nu^6+10 \nu^8) \nu,3628800) v_1^2)+\cr
&q_1^2 (\ffc(0,
1) v_1+\ffc(-(2 \nu+1) (5 \nu^8+20 \nu^7-40 \nu^6-190 \nu^5+158 \nu^4+656 \nu^3-378 \nu^2-741 \nu
+381),7257600) v_1^2)+...,
}}
The prediction of \OV\ is that the above expansions in $q_1$ and
$v_1$ are integral when rewritten in the form \mcf\ for 
any choice of an integer $\nu\in \ZZ$. The 
invariants $n^{h=1}_{g,n,w}$ for general framing $\nu$ are 
collected in the appendix. As expected on general grounds, 
these invariants are independent 
of the framing $\nu$ for $w=1$, except for a phase factor $\eps=(-1)^\nu$.
For $w>1$ the $n_{g,n,w}$ are polynomials
$p_\delta(\nu,\eps)$ of degree $\delta=2g+w-1$ in $\nu$. 
A closed proof of integrality of all the 
polynomials $p_\delta(\nu,\eps)$ would be formidable.
We contented ourselves to a verification of 
the integrality of the $p^{h=1}_\delta(\nu,\eps)$ for 
a large number of values of $\nu$. For framing $\nu=0$,
the only nonzero invariants are $n_{0,0,1}=-1$ and $n_{0,1,1}=1$\foot{Here
and in the following tables there is a factor $(-1)^h$ 
in our conventions relative to \AVi\GZ.}. The result 
for framing $\nu=\pm 1$ is:
\def\ss#1{{\scriptstyle #1}}
\def\ps{{\phantom{\{}}}
$$
\vbox{\offinterlineskip\tabskip=-3pt\halign{
#
&\hfil~$\ss{#}$&\hfil~$\ss{#}$&\hfil~$\ss{#}$
&\hfil~$\ss{#}$&\hfil~$\ss{#}$&\hfil~$\ss{#}$
&\hfil~$\ss{#}$&\hfil~$\ss{#}\ps$
\cr
\noalign{\hrule}
&&&&&&&&g=0\cr
\noalign{\hrule}
&1& -1& 1& -2& 5& -13& 35& -100\cr 
&-1& 1& -2& 5& -14& 42& -132& 429\cr &
0& 0& 1& -4& 14& -52& 198& -752\cr 
&0& 0& 0& 1& -6& 31& -150& 693\cr 
\noalign{\hrule}
&&&&&&&&g=1\cr
\noalign{\hrule}
&0& 0& -1& 6& -32& 156& -718& 3220\cr 
&0& 0& 1& -10& 70& -420& 2310& -12012\cr 
&0& 0& 0& 4& -49& 406& -2838& 17840\cr 
&0& 0& 0& 0& 11& -166& 1650& -13398\cr 
\noalign{\hrule}
&&&&&&&&g=2\cr
\noalign{\hrule}
&0& 0& 0& -5& 76& -772& 6356& -45990\cr 
&0& 0& 0& 6& -133& 1743& -17556& 150150\cr 
&0& 0& 0& -1& 63& -1300& 17655& -189260\cr 
&0& 0& 0& 0& -6& 351& -7785& 115269\cr
\noalign{\hrule}
&&&&&&&&g=3\cr
\noalign{\hrule}
&0& 0& 0& 1& -85& 2059& -32037& 386484\cr 
&0& 0& 0& -1& 121& -3926& 76571& -1111682\cr 
&0& 0& 0& 0& -37&2241& -63063& 1191808\cr 
&0& 0& 0& 0& 1& -382& 20825& -586146\cr 
\noalign{\hrule}
&&&&&&&&g=4\cr
\noalign{\hrule}
&0& 0& 0& 0& 45& -3225& 102243& -2138540\cr
&0& 0& 0& 0& -55& 5291& -213785& 5460026\cr 
&0& 0& 0& 0& 10& -2297& 144430& -4992704\cr 
&0& 0& 0& 0& 0& 232& -35221& 1974995\cr 
\noalign{\hrule}}}
%
\hskip 30pt
\vbox{\offinterlineskip\tabskip=-3pt\halign{
#
&\hfil~$\ss{#}\ps$&\hfil~$\ss{#}$&\hfil~$\ss{#}$
&\hfil~$\ss{#}$&\hfil~$\ss{#}$&\hfil~$\ss{#}$
&\hfil~$\ss{#}$&\hfil~$\ss{#}\ps$
\cr
\noalign{\hrule}
&&&&&&&&g=0\cr
\noalign{\hrule}
 &1& 0& 0& 0& 0& 0& 0& 0\cr &-1& -1& -1& -1& -1& -1& -1& -1\cr &0& 1& 2& 4
& 6& 9& 12& 16\cr &0& 0& -1& -5& -14& -31& -60& -105\cr 
\noalign{\hrule}
&&&&&&&&g=1\cr
\noalign{\hrule}
&0& 0& 0& 0& 0& 0& 0& 0\cr &0& 0& 0& 0& 0& 0& 0& 0\cr &0& 0& -1&
-4& -11& -24& -46& -80\cr &0& 0& 1& 10& 49& 166& 450& 1050\cr 
\noalign{\hrule}
&&&&&&&&g=2\cr
\noalign{\hrule}
&0& 0& 0& 0& 0& 0& 0& 0\cr &0& 0& 0& 0& 0& 0& 0&
0\cr &0& 0& 0& 1& 6& 22& 62& 148\cr &0& 0& 0& -6& -63& -351& -1392& -4431\cr 
\noalign{\hrule}
&&&&&&&&g=3\cr
\noalign{\hrule}
&0& 0& 0& 0& 0& 0& 0& 0\cr &0&
0& 0& 0& 0& 0& 0& 0\cr &0& 0& 0& 0& -1& -8& -37& -128\cr &0& 0& 0& 1& 37& 382& 
2333& 10424\cr
\noalign{\hrule}
&&&&&&&&g=4\cr
\noalign{\hrule}
&0& 0& 0& 0& 0
& 0& 0& 0\cr &0& 0& 0& 0& 0& 0& 0& 0\cr &0& 0& 0& 0& 0& 1& 10& 56\cr 
&0& 0& 0& 0&-10& -232& -2343& -15233\cr  
\noalign{\hrule}}}
$$
\vbox{\leftskip 2pc\rightskip 2pc
\noindent{\ninepoint
{\bf Table 1\rrr:} Low degree Ooguri--Vafa (OV) invariants $n^{h=1}_{g,n,w}$ 
for framing $\nu=1$ (left) and $\nu=-1$ (right). 
The degree $n\geq 0$ (winding $w\geq 1$) corresponds to the 
vertical (horizontal) direction.}}

\noi
The integrality of the above invariants provides a highly 
non-trivial verification on the M-theory predictions of \OV\LMV, 
and the localization methods proposed in refs.\LS\KL\GZ. 

\goodbreak
\subsubsec{4.2.2.\rrr
The gauge kinetic function and higher genus generalizations}
The other holomorphic coupling in the standard $\cx N=1$ supergravity
is the gauge kinetic $f$-function related to the partition function
$\cx F_{0,2}$. Higher genus partition functions contribute to 
$f$ with a coefficient proportional to the $2g$-th power of the 
vev of the graviphoton field strength.

To compute $\cx F_{g,2}$ we 
consider the sum over graphs with two distinguished legs which 
computes the Gromov-Witten invariants $F_{g,n;w_1,w_2}$, associated
to world-sheets with two boundaries landing on two distinguished
parallel branes. As discussed in sect. 2\rrr these GW invariants 
are the coefficients of the monomial $v_1^{w_1}v_2^{w_2}$. 
Restricting to the terms with 
$w_1\leq w_2$, the first terms in the instanton expansion of the
partition function $\cx F_{0,2}$ are 
\eqn\cfpfii{\eqalign{
{\cal F}^*_{0,2}=
&\ffc(-n (-1+n),2) v_1 v_2+\ffc(n (-1+n) (-1+2 n),3) v_1 v_2^2+\ffc(-n (
-1+n) (-1+2 n)^2,4) v_1^2 v_2^2+\cr
&q_1 (\ffc(n^2,1) v_1 v_2+\ffc(-n^2 (-1+2 n),1) 
v_1 v_2^2+\ffc(n^2 (-1+2 n)^2,1) v_1^2 v_2^2)+\cr
&q_1^2 (\ffc(-(1+n) n,2) 
v_1 v_2+\ffc(n^2 (2 n+1),1) v_1 v_2^2+\ffc(-n^2 (12 n^2-1),2) v_1^2 v_2^2)+\cr
&q_1^
3 (\ffc(-n (1+n) (2 n+1),3) v_1 v_2^2+\ffc(n^2 (2 n+1)^2
,1) v_1^2 v_2^2)+\cr
&q_1^4 (\ffc(-n (1+n) (
2 n+1)^2,4) v_1^2 v_2^2)+...\,  ,\cr
}}
where the star is to remind that there are other terms
in $\cx F_{0,2}$ following from the general form \gpf.
The partition function $\cx F_{0,2}$ for a single brane is 
obtained by setting $v_1=v_2$ in the above. Similar
statements apply to the partition functions $\cx F_{g,2}$ 
for $g>0$, which can be found in the appendix.

Taking into account the multi-coverings leads to 
the integral invariants $n_{g,\be}$. We restrict to 
quote the low degree invariants for two particular choices of framings
and refer to the appendix for more detailed results. 
The partition functions $\cx F_{g,2}$ are identically zero 
for zero framing
$$
\nu=0:\qquad n^{h=2}_{g,n,w_1,w_2}=0,\qquad \forall\, g,n,w_1,w_2.
$$
This extends the findings of 
\LS\KL\ for degree zero, $n=0$ to all other degrees.
For framing $\nu=-1$ we find 
\vfil
\goodbreak

\def\ss#1{{\scriptstyle #1}}
\def\ps{{\phantom{\bigoplus}}}
$$
\vbox{\offinterlineskip\tabskip=-4pt\halign{
#
&\hfil~$\ss{#}$~&\hfil~$\ss{#}$~&\hfil~$\ss{#}$~&\hfil~$\ss{#}\ps$~\vrule
&\hfil~$\,\ss{#}$~&\hfil~$\ss{#}$~&\hfil~$\ss{#}$~&\hfil~$\ss{#}\ps$~\vrule
&\hfil~$\,\ss{#}$~&\hfil~$\ss{#}$~&\hfil~$\ss{#}$~&\hfil~$\ss{#}\ps$~\vrule
&\hfil~$\,\ss{#}$~&\hfil~$\ss{#}$~&\hfil~$\ss{#}$~&\hfil~$\ss{#}\ps$~
\cr
\noalign{\hrule}
&&&&&&&&&&&&&&&&g=0\cr
\noalign{\hrule}
&0& 0& 0& 0&\hskip3pt 1
& 1& 1& 1& -1& -3& -6& -10& 0& 2& 10& 30\cr &0& 0& 0& 0&
1& 1& 1& 1& -3& -6& -9& -14& 2& 9& 26& 60\cr &0& 0& 0& 0& 1& 1& 1& 1& -6& -9& -\
13& -18& 10& 26& 57& 112\cr &0& 0& 0& 0& 1& 1& 1& 1& -10& -14& -18& -24& 30& 60
& 112& 195\cr
\noalign{\hrule}
&&&&&&&&&&&&&&&&g=1\cr
\noalign{\hrule}
&0& 0& 0& 0& 0& 0& 0& 0& 0& 1& 5& 15& 0& -1& -15& -85\cr &0& 0& 0& 0& 0
& 0& 0& 0& 1& 4& 12& 28& -1& -12& -65& -240\cr &0& 0& 0& 0& 0& 0& 0& 0& 5& 12&
26& 51& -15& -65& -220& -616\cr &0& 0& 0& 0& 0& 0& 0& 0& 15& 28& 51& 88& -85& -\
240& -616& -1430\cr
\noalign{\hrule}
&&&&&&&&&&&&&&&&g=2\cr
\noalign{\hrule}
&0& 0& 0& 0& 0& 0& 0& 0& 0& 0& -1& -7& 0& 0& 7& 91\cr &0& 0& 0& 0& 0& 0
& 0& 0& 0& -1& -6& -23& 0& 6& 70& 427\cr &0& 0& 0& 0& 0& 0& 0& 0& -1& -6& -22&
-63& 7& 70& 395& 1622\cr &0& 0& 0& 0& 0& 0& 0& 0& -7& -23& -63& -150& 91& 427&
1622& 5159\cr
\noalign{\hrule}
&&&&&&&&&&&&&&&&g=3\cr
\noalign{\hrule}
&0& 0& 0& 0& 0& 0& 0& 0& 0& 0& 0& 1& 0& 0& -1& -46\cr &0& 0& 0& 0& 0& 0
& 0& 0& 0& 0& 1& 8& 0& -1& -38& -421\cr &0& 0& 0& 0& 0& 0& 0& 0& 0& 1& 8& 37& -\
1& -38& -398& -2501\cr &0& 0& 0& 0& 0& 0& 0& 0& 1& 8& 37& 128& -46& -421& -2501
& -11192\cr
\noalign{\hrule}
&&&&&&&&&&&&&&&&g=4\cr
\noalign{\hrule}
&0& 0& 0& 0& 0& 0& 0& 0& 0& 0& 0& 0&  &  & w_1\,\to &  \cr &0& 0& 0& 0&
0& 0& 0& 0& 0& 0& 0& -1&  & w_2 &  &  \cr &0& 0& 0& 0& 0& 0& 0& 0& 0& 0& -\
1& -10&  & \downarrow &  &  \cr &0& 0& 0& 0& 0& 0& 0& 0& 0& -1& -10& -56&  & 
 &  &  \cr
\noalign{\hrule}}}
$$
\vbox{\leftskip 2pc\rightskip 2pc
\noindent{\ninepoint
{\bf Table 2\rrr:} Low degree OV invariants $n^{h=2}_{g,n,w_1,w_2}$ 
for degree $n\geq 0$ (increasing to the right) and framing $\nu=-1$.\br\  }}
Again these and the other results reported in the appendix are in impressive
agreement with the integrality prediction.

\subsubsec{4.2.3.\rrr More boundaries and general structures}
It is interesting to study also the general structure of the 
partition functions with more boundaries, which describe
higher dimension operators in the D-brane gauge theory
and couplings involving powers of the graviphoton superfield\foot{See 
\vafaln\ for an interpretation
of these couplings in the D-brane gauge theory.}.
The computation of the graph sums with $h$ distinguished 
legs results in the following expression for the 
leading terms in the expansion of the genus zero
partition functions $\cx F_{0,h}$:
$$\eqalign{
\cx F^*_{0,3} = 
&\ffc(-\nu^3 (-2+3 \nu),1) q_1 v_1 v_2 v_3+\ffc(\nu^3 (-1+2 \nu) (-3+4 \nu),1) q_1
 v_1 v_2 v_3^2+\ffc(-\nu^3 (-4+5 \nu) (-1+2 \nu)^2,1) q_1 v_1 v_2^2 v_3^2+\cr
&\ffc(\nu^3 (-5+6 \nu) (-1+2 \nu)^3,1) q_1 v_1^2 v_2^2 v_3^2+\ffc((2+3 \nu) 
\nu^3,1) q_1^2 v_1 v_2 v_3+\ffc(-2 \nu^3 (6 \nu^2-1),1) q_1^2 v_1 v_2 v_3^2+\cr
&\ffc(2 \nu^3 (-1+2 \nu) (10 \nu^2-\nu-1),1) q_1^2 v_1 v_2^2 v_3^2+\ffc(-(-1+2 \nu)^2 \nu^3 (30 \nu^2-5 \nu-2),1) q_1^2 v_1^2 
v_2^2 v_3^2+... \cr
\cx F^*_{1,3} = 
&\ffc((10-9 \nu-24 \nu^2+24 \nu^3) \nu^3,24) q_1 v_1 v_2 v_3+
\ffc(-(15-65 \nu^2+52 \nu^3) \nu^3 (-1+2 \nu),12) q_1 v_1 v_2 v_3^2+\cr
&\ffc((68+51 \nu-392 \nu^2+280 \nu^3) \nu^3 (-1+2 \nu)^2,24) q_1 v_1 v_2^2 v_3^2+\cr
&\ffc(-5 (13+18 \nu-90 \nu^2+60 \nu^3) \nu^3 (-1+2 \nu)^3,12) q_1 v_1^2 v_2^2 v_3^2+...\cr
\cx F^*_{0,4} = 
&\ffc(\nu^4 (-3+4 \nu)^2,1) q_1 v_1 v_2 v_3 v_4+\ffc(-\nu^4 (-4+5 \nu)^2 (-1+2 \nu),1) q_1 v_1 v_2 v_3 v_4^2+\cr
&\ffc(\nu^4 (-5+6 \nu)^2 (-1+2 \nu)^2,1) q_1 v_1 v_2 v_3^2 v_4^
2+\ffc(-\nu^4 (-6+7 \nu)^2 (-1+2 \nu)^3,1) q_1 v_1 v_2^2 v_3^2 v_4^2+\cr
&\ffc(\nu^4 (-7+8 \nu)^2 (-1+2 \nu)^4,1) q_1 v_1^2 v_2^2 v_3^2 v_4^2+...\cr
}$$
The polynomial dependence of the GW invariants on the
framing $\nu$ is described by the simple relation\foot{
The degree in $\nu$ coincides with 
the dimension of the moduli space $\cx M_g^{rel}(\IP^1,\be)$
defined in \LS.}
\eqn\genform{
n_{g, n, \vx w} = p(\nu,\eps),\qquad \rmx{deg_\nu}(p)=2g-2+h+w,}
with $w=\sum_\al w_\al$.
The above expansions vanish again at $\nu=0$, as do 
all partition functions $\cx F_{g,h}$ considered so far, except
for the two non-zero instantons contributing to $\cx F_{0,1}$.
As for non-zero framing, subtracting the multi-cover contributions 
one obtains the invariants defined in \OV;  for all classes
and framings we have considered they are integral. 
Some low degree invariants for framing $\nu=-1$ are:
\def\ss#1{{\scriptstyle #1}}
\def\ps{{\phantom{\int}}}
$$
\vbox{\offinterlineskip\tabskip=-1pt\halign{
#
&\hfil~$\ss{#}\ps$&\hfil~$\ss{#}$&\hfil~$\ss{#}$&\hfil~$\ss{#}$
&\hfil~$\ss{#}$&\hfil~$\ss{#}$&\hfil~$\ss{#}$&\hfil~$\ss{#}$
&\hfil~$\ss{#}$&\hfil~$\ss{#}$
\cr
\noalign{\hrule}
&&&& &&&&& h=3 & g=0\ps\cr
\noalign{\hrule}
&0& 0& 0& 0& 0& 0& 0& 0& 0& 0\cr &-1& -1& -1& -1& -1& -1& -1& -1& -1& -1
\cr &5& 10& 17& 16& 24& 33& 24& 32& 42& 53\cr &-4& -21& -66& -60& -141& -276& -\
130& -258& -453& -729\cr &0& 12& 90& 81& 328& 964& 292& 881& 2136& 4502\cr
\noalign{\hrule}
&&&& &&&&& h=3  & g=1\ps\cr
\noalign{\hrule}
&0& 0& 0& 0& 0& 0& 0& 0& 0& 0\cr &0& 0& 0& 0& 0& 0& 0& 0& 0& 0\cr &-1& -6
& -20& -17& -41& -80& -36& -73& -128& -205\cr &1& 22& 145& 119& 462& 1302& 394&
1148& 2718& 5598\cr &0& -16& -275& -222& -1536& -6602& -1282& -5692& -18792& -\
50951\cr
\noalign{\hrule}}}
$$
\vbox{\leftskip 2pc\rightskip 2pc
\noindent{\ninepoint
{\bf Table 3\rrr:} Low degree OV invariants $n^{h=3}_{g,n,\vx w}$ 
for framing $\nu=-1$. The degree $n\geq 0$ corresponds to the 
vertical direction and the windings $w_a\leq 3$ to the horizontal
direction with 3-tuples sorted in increasing numerical order.}}

$$
\vbox{\offinterlineskip\tabskip=-4pt\halign{
#
&\hfil~$\ss{#}\ps$&\hfil~$\ss{#}$&\hfil~$\ss{#}$&\hfil~$\ss{#}$
&\hfil~$\ss{#}$&\hfil~$\ss{#}$&\hfil~$\ss{#}$&\hfil~$\ss{#}$
&\hfil~$\ss{#}\ps$&\hfil~$\ss{#}$&\hfil~$\ss{#}$&\hfil~$\ss{#}$
&\hfil~$\ss{#}$&\hfil~$\ss{#}$&\hfil~$\ss{#}$
\cr
\noalign{\hrule}
&&&& &&&& &&&& && h=4 & g=0\ps\cr
\noalign{\hrule}
&0& 0& 0& 0& 0& 0& 0& 0& 0& 0& 0& 0& 0& 0& 0\cr &1& 1& 1& 1& 1& 1& 1& 1&
1& 1& 1& 1& 1& 1& 1\cr &-18& -30& -46& -44& -62& -82& -60& -80& -102& -126& -80
& -100& -124& -150& -178\cr &49& 152& 375& 346& 708& 1269& 662& 1200& 1980& 
3051& 1135& 1890& 2934& 4320& 6102\cr &-32& -243& -1060& -960& -2993& -7512& -\
2736& -6936& -15131& -29580& -6368& -14104& -27808& -50409& -85584\cr
\noalign{\hrule}}}
$$
\vbox{\leftskip 2pc\rightskip 2pc
\noindent{\ninepoint
{\bf Table 4\rrr:} Low degree OV invariants $n^{h=4}_{g,n,\vx w}$ 
for framing $\nu=-1$. The degree $n\geq 0$ corresponds to the 
vertical direction and the windings $w_a\leq 3$ to the horizontal
direction with 4-tuples sorted in increasing numerical order.}}

\subsec{Case II: The canonical bundle $K_{\IP^2}$}
A similar computation leads to the open string partition
functions $\cx F_{g,h}$ for the two phases of D-branes on the 
second one moduli case, the canonical bundle on $\IP^2$.
One phase was already considered in the 
paper of Graber and Zaslow \GZ\ and the one boundary 
partition functions $F_{g,1}$ contributing to
the superpotential were discussed in detail.
The extension of these computation to other world-sheet
topologies and D-brane phases is a 
straightforward elaboration on their work.
Moreover the arguments and computations are similar to those in the 
discussion of $\cf$ and we can thus be brief in the following.

The equivariant class of the pull-back $i^*\phi$ computes to \YZ
\eqn\intpii{
\prod_{e\in E} \prod_{a=1}^{3d-1}(\La_i+\fc{a}{d}(\la_i-\la_j))
\times 
\prod_{v\in V} \big(\La_i^{val-1}\, P_g(\La_i)\big)}
where $\La_i=\sum_{a=1}^3\la_a-3\la_i$ serves as 
one possible linearization. 
There are essentially again two distinguished phases for the
D-branes, 
depending on whether the center of $\CC$ lands on 
the compact divisor $\IP^2$ $(I)$ or not $(II)$.
The primitive disc $D:\, z_a=z_b=0,\, |z_D|^2\leq c$ 
and the choice of a direction in the
charge lattice of $U(1)^2$ for these two phases are:
\eqn\defwkb{
\vbox{\offinterlineskip\tabskip=-4pt\halign{
\strut
~$#$\hfil~\ &~$#$\hfil~\ &~$#$\hfil~\ &~$#$\hfil~\qquad &~$#$\hfil~\cr
I:&z_D=z_1/z_3&,z_a=z_2/z_3&,z_b=z_4z_3^3&\nu=\fc{\la_2-\la_3}{\la_1-\la_3},\cr\cr
II:&z_D=z_4z_3^3&,z_a=z_1/z_3&,z_b=z_2/z_3&
\nu=\fc{\la_1-\la_3}{-\La_3},\cr
}}}
The open string integrand for the integrals in \abo\ is
given by the product \intcs$\cdot$\intL$\cdot$\intpii.
Results for the partition functions and OV invariants 
are collected in the appendix.

\newsec{The A-model and local mirror symmetry on a \CY 4-fold}
It is for several reasons interesting to study the topological
closed string on \CY 4-folds $Z$. The type II string
compactified on $Z$ has the same number of supercharges
as the type II theory on a \CY 3-fold with supersymmetric branes.
A much stronger connection exists for the class of open string
theories considered above: there is a proposal for an open/closed string
duality \osi\ which associates to a D-brane geometry on a chosen \CY
3-fold $X$ a closed string theory on a specific \CY 4-fold $Z$
such that the superpotential of the 2d closed string is identical 
to that of the four-dimensional $\cx N=1$ supersymmetric
open string theory. The superpotential of the two-dimensional
closed string is proportional to the genus zero partition function
and the equality of the superpotentials amounts to the
relation $\cx F_{0,0}(Z)=\cx F_{0,1}(X)$, which has been shown
on the level of the full world-sheet instanton expansion. This is 
a very strong indication for a deeper relation between
the two theories and possibly a true
string duality, perhaps involving F-theory compactified on $Z$.
It would be interesting to study this relation further on the level of
other world-sheet genera.

Whereas the genus zero partition function can be computed
by mirror symmetry for \CY 4-folds \gmp\ffs, and can be interpreted
in terms of ``numbers'' of holomorphic curves in $Z$,  
the higher genus case has not been considered so far. 
In the following we outline the use of localization techniques
for the computation of the genus zero and one partition functions 
for the basic non-compact \CY 4-fold $\kbff$ with $h^{1,1}=1$. 
It would be interesting to extend these computations to the
moderately more complicated \CY 4-folds relevant for the
open/closed string duality of \osi.

\subsubsec{Genus zero computation and comparison with local mirror symmetry}
\noi
The basic genus zero GW invariants on a \CY $d$-fold $Z$ 
are defined as the coefficients in the expansion of the topological 3-point 
amplitude
\eqn\tcf{
\langle \cx O^{(1)}\cx O^{(1)} \cx O^{(d-2)} \rangle,
}
where the operators $\cx O^{(k)}$ are associated to 
elements $\ga^{(d-k)}$ in $H_*(Z)$ of complex 
codimension $k$ \Wtsm. The new aspect
of the 4-fold is that the last operator in \tcf\ has codimension
two in $Z$ and thus the generic holomorphic curve does not 
intersect the class $\ga^{(d-k)}$. The genus zero
partition function for the basic GW invariants 
has the formal instanton expansion \KM
$$
\cx F_{0} =
\sum_{\be \neq 0}\, \sum_{a} \tx \ga^{(2)}_a F_{0,a,\be}\, q(\be)
$$
where $\be$ denotes again a class in $H_2(Z,\ZZ)$,
$q(\be)$ its complexified K\"ahler volume and moreover
the last sum runs over a basis of formal coordinates 
$\tx \ga_a^{(2)}$ dual to the homology classes of codimension two.
The genus zero GW invariants $F_{0,a,\be}$ can be 
interpreted as the ``number'' of holomorphic curves that
intersect the 4-cycle $\ga_a^{(2)}$. The multi-cover structure
that exposes the integral invariants $n_{0,a,\be}$ associated to 
the virtual Euler number of moduli spaces of such curves is \ffs\gmp
\eqn\mcfff{
F_{0,a,\be} = \sum_{\be'\cdot d = \be } d^{-2}\, n_{0,a,\be'}.
}
Note that the multi-cover structure of the closed string
theory compactification to two dimensions is identical to that
of the open string theory \mcf, at $(g,h)=(0,1)$.\foot{In fact also the 
open string result is derived in \OV\ using a two-dimensional
variant of the open string background.} For
the relation between the open/closed string duals considered
in \osi\ this amounts to the statement that the integral 
open string disc invariants $n^{h=1}_{0,\be}$ map 1-1
to the closed string integral invariants $n_{0,a,\hx \be}$,
after an appropriate identification of the classes in 
$\be\in H^{rel}_2(X,L;\ZZ)$ with classes in $\hx \be\in H(Z,\ZZ)$.

What makes it necessary to impose an condition on the 
location of the curves in the A-model computation is the dimension of
the moduli space of curves in general position, which is too large
to find an appropriate form on it.
To obtain a moduli space of the right dimension with 
a top form that can be integrated over it, one may add one
marked point $x$ on the domain curve $\Si$ and requires the image
$f(x)$ in $Z$ to lie within $\ga_a^{(2)}$. The computation of
the weights of the Euler class of the obstruction bundle and
the virtual normal bundle leads to the result 
\eqn\intcsff{
\eqalign{
&\prod_{e\in E} \big[ \fc{(-)^d}{d!^2}(\fc{d}{\la_{i}-\la_{j}})^{2d}
\prod_{k\neq i,j \atop a=0}^{a=d} (\fc{a}{d}\la_{i}+
\fc{d-a}{d}\la_{j}-\la_k)^{-1}
\prod_{a=1}^{4d-1}(\La_i+\fc{a}{d}(\la_i-\la_j))\big] \times\cr
&\prod_{v\in V}\big[
\big(\La_i\prod_{k\neq i}(\la_i-\la_k)\big)^{val-1}
(\la_i)^{2p}\times \{(\sum_{F\ni v} 
w_F^{-1})^{val+p-3}\prod_{F\ni v} w_F^{-1}\}
\ \big],
}}
where $p(v)$ is the number of marked points at a vertex $p$
and the indices of the weights run over the four fixed points $p_i$  
on $\IP^3$.
Moreover, in a particular linearization, $\La_i=\sum_{k=1}^4 \la_k -\la_i$.
The graph sum \defkd\ leads to the GW invariants collected
in the genus zero partition function 
\eqn\pfff{
\cx F_0 = {\sevenrm
-20 q_1-825 q_1^2-\fc{612560}{9} q_1^3-\fc{29946585}{4} q_1^4
-\fc{4825194504}{5} q_1^5-\fc{412709577260}{3} q_1^6 +}...}
Here $q_1$ is the exponentiated K\"ahler volume of 
the single primitive class $\be\in H_2(Z,\ZZ)$ and 
we have omitted the label for the single class $\ga^{(2)}$ in  $H_4(Z)$.
From the genus zero multi-cover formula \mcfff\ for the 4-fold 
one obtains the invariants $n_{0,n\cdot \be}$
$$
-20,\  -820,\  -68060,\  -7486440,\  -965038900,\  -137569841980,\  ...
$$
which are indeed all integers.

The above result agrees with the one obtained from 
mirror symmetry for \CY 4-folds.
The GW invariants associated to the
compact divisor $\IP^3$ in $\kbff$ can be
computed from an application of local mirror symmetry as in \amv.
The non-compact toric 4-fold $Z$ may be defined as the 
gauged $U(1)$ linear sigma model with five matter fields 
of charges $l=(-4,1,1,1,1)$. The periods of the local mirror manifold
are the solutions to the (GKZ) system of differential equations
associated to the vector $l$:
\def\thz{\theta}
$$
\cx L = \thz^4-8z \thz (4 \thz+1) (2 \thz+1) (4 \thz+3),
$$
where $z$ is the single complex structure modulus of the mirror and
$\theta = z\fc{d}{dz}$. Apart from the constant solution $\om_0=$const., 
the solutions $\om_k$ of the differential operator $\cx L$ 
are of the form $\om_k=ln(z)^k+S_k(z),\, 
k=1,2,3,$ where $S_k(z)$ are power series in $z$.
The single logarithmic solution describes the mirror map
$$
2\pi i t=\om_1(z)=\ln(z)+24 z+1260 z^2+123200 z^3+15765750 z^4+...$$ 
relating the K\"ahler modulus $t$
of $\kbff$ and the complex structure 
modulus $z$ of its mirror. The partition function 
$\cx F_0$ is given by the double logarithmic solution $\om_2(z)$ \ffs.
Inverting the mirror map and inserting the result $z(t)$ into
$\om_2$ leads to a power series in $q_1=exp(2\pi i t)$ that agrees with the
result from the localization computation \pfff.

\subsubsec{Genus one}
\noi
At genus one, the moduli space of generic curves is 
of the right dimension for any dimension $d$ of the \CY\
target space and the computation involves maps without
extra marked points. The genus one integrand is obtained from 
that at genus zero by replacing the expression 
$\{(\sum_{F\ni v} w_F^{-1})^{val+p-3}\prod_{F\ni v} w_F^{-1}\}$ 
in the last line of \intcsff\ by the class 
\eqn\intcdffii{\{\fc{P_1(\La_i)\cdot \prod_{i\neq k} P_1(\la_i-\la_k)}
{\prod_{F\ni v} w_F-\psi_F}\}}
The graph sum \defkd\ leads to the following expansion of the
genus one partition function:
\eqn\pfffii{
\cx F_1 = -\fc{25}{3}q_1-\fc{2425}{6}q_1^2-\fc{204700}{9}q_1^3
+\fc{688375}{12}q_1^4+{492685322}q_1^5+\fc{1433052348850}{9}q_1^6+...}
It would be interesting to relate this partition function to the
number of elliptic curves in $Z$. 

\newsec{Discussion}
The improved techniques for $\cx N=1$ open strings 
developped over the last years have put the 
study of non-perturbative properties
of $\cx N=1$ string vacua on a new level. 
Important non-perturbative aspects, such as 
supersymmetry breaking and a lift 
of the vacuum degeneracy, have been discussed 
in a plethora of papers in the past, mainly
based on qualitative arguments and some 
reasonable working hypotheses. It will be interesting to 
check some of these ideas now and to redo the analysis 
with the new quantitative techniques that emerged
from the D-brane techniques. 

The computations in the previous sections provide
the instanton expansions of the basic holomorphic
couplings $W(\phi)$ and $f(\phi)$ in the low energy 
string effective action and are thus a modest
first step into this direction. One of 
the primary motivations of the computation of
the explicit instanton expansions in this paper
has been the wish to use them as a starting
point for the development of methods that govern
also the global structure on the  $\cx N=1$ ``moduli space''.
This is the parameter space of scalar vev's which are
flat directions in the perturbative sense. The
instanton expansions do not lead immediately to information 
about the global behavior of the couplings.
One promising approach to uncover the principles
that govern this global structure would be  a generalization
of mirror symmetry arguments along the lines of \bcov\Cgms\AVi.
Another possible strategy would be to derive
a system of differential equations satisfied by the 
instanton expansions for $\cx F_{g,h}$, such as \cfpfi\ and \cfpfii\ 
for the conifold. For the superpotential $W=\cx F_{0,1}$, 
such equations are known \osi\osii.
\vskip 20pt

\noi
{\bf Acknowledgments:}\br 
I would like to thank Eric Zaslow for many explanations on the work \GZ\
and Carel Faber for providing the Maple implementation of the work \Fab.
I am grateful to Wolfgang Lerche for valuable discussions.

\goodbreak
\appendix{A}{Integral invariants for general framings}
\def\ffc(#1,#2){\ifnum #2=1 {\scriptstyle #1} 
\else {\scriptstyle {{1}\over{#2}}(#1}) \fi}
In the following we restrict to describe the low degree
OV invariants for general framing in 
the phase of D-branes on $\cf$ studied in sect. 4\rrr.
An extended appendix including results for other phases
and D-branes on the \CY 3-fold $\kb$ can be found at 
\app.

The results will be presented in terms of generating functions $\cx A_{g,h}$
similar to the partition functions $\cx F_{g,h}$, but with 
the coefficients being the integral OV invariants $n_{g,h,\be}$ instead of the
fractional GW invariants $F_{g,h,\be}$.
The first terms in the
generating functions for world-sheets with $h=1$ boundary
are:
$$\eqalign{
{\cal A}_{0,1}=
&\ffc(-\eps,1) v_1+\ffc(-2 \nu-1+\eps,4) v_1^2+\ffc(-\eps \nu (1+\nu),2) v_1^3+
\ffc(-\nu (1+2 \nu) (1+\nu),3) v_1^4+\hskip 200pt\cr
&q_1 (\ffc(\eps,1) v_1+\ffc(\nu,1) v_1^2+\ffc(\eps
 \nu (1+3 \nu),2) v_1^3+\ffc(\nu (1+2 \nu) (1+4 \nu),3) v_1^4)+\cr
&q_1^2 (\ffc(-2 \nu+1-\eps,4) v_1^2+\ffc(-\eps \nu (-1+3 \nu),2) v_1^3+\ffc(-4 \nu^3,1) 
v_1^4)+\cr
&q_1^3 (\ffc(\eps \nu (-1+\nu),2) v_1^3+\ffc(
(-1+4 \nu) (-1+2 \nu) \nu,3) v_1^4)+ ...\cr
}$$
$$\eqalign{
&{\cal A}_{1,1}=\cr
&\ffc(-4 \nu-3+6 \nu^2+4 \nu^3+3 \eps,48) v_1^2+\ffc(\eps \nu (1+\nu) 
(3 \nu^2+3 \nu-2),8) v_1^3+\ffc(\nu (1+2 \nu) (1+\nu) (2 \nu^2+2 \nu-1),3) v_1^4+\cr
&q_1 (\ffc(-\nu (-1+\nu) (1+\nu),6) v_1^2+\ffc(-\eps \nu (1+3 \nu) (-2+3 \nu) (1+
\nu),8) v_1^3+\ffc(-\nu (1+4 \nu) (1+2 \nu) (-1+2 \nu) (1+\nu),3) v_1^4)+\cr
&q_1^2 (\ffc(-4 \nu+3-6 \nu^2+4 \nu^3-3 \eps,48) v_1^2+\ffc(\eps \nu (-1+\nu) (3 \nu+2) (-1
+3 \nu),8) v_1^3+\ffc(4 \nu^3 (-1+2 \nu^2),1) v_1^4)+\cr
&q_1^3 (\ffc(-\eps \nu (-1+\nu) (3 \nu^2-3 \nu-2),8) v_1^3+\ffc(-\nu (-1+\nu) (1+2 \nu)
 (-1+2 \nu) (-1+4 \nu),3) v_1^4)+...\cr
}$$
$$\eqalign{
&{\cal A}_{2,1}=\cr
&\ffc(-16 \nu-15+40 \nu^2+20 \nu^3-10 \nu^4-4 \nu^5+15 \eps,960) 
v_1^2+\ffc(-\eps \nu (-1+\nu) (-1+3 \nu) (3 \nu+4) (2+\nu) (1+\nu),80) v_1^3+\cr
&q_1 (\ffc(\nu (-1+\nu) (\nu-2) (2+\nu) (1+\nu),120) v_1^2+\ffc(\eps \nu (-1+\nu) (3 \nu+4) (1+
3 \nu) (-2+3 \nu) (1+\nu),80) v_1^3)+\cr
&q_1^2 (\ffc(-16 \nu+15-40 \nu^2+20
 \nu^3+10 \nu^4-4 \nu^5-15 \eps,960) v_1^2+\ffc(-\eps \nu (-1+\nu) (3 \nu+2) (-1+3 \nu) (3 \nu-4
) (1+\nu),80) v_1^3)+\cr
&q_1^3 (\ffc(\eps \nu (-1+\nu) 
(\nu-2) (3 \nu-4) (1+3 \nu) (1+\nu),80) v_1^3)+...\cr
}$$
$$\eqalign{
&{\cal A}_{3,1}=\cr
&\ffc(28 \nu^6+8 \nu^7-288 \nu-315+952 \nu^2+392 \nu^3-350 \nu^4-112 \nu
^5+315 \eps,80640) v_1^2+\hskip 200pt\cr
&q_1 (\ffc(-\nu (-1+\nu) (\nu-2) (-3+\nu) (3+\nu)
 (2+\nu) (1+\nu),5040) v_1^2)+\cr
&q_1^2 (\ffc(-28 \nu^6+8 \nu^7-288 \nu+315
-952 \nu^2+392 \nu^3+350 \nu^4-112 \nu^5-315 \eps,80640) v_1^2)+...\cr
}$$
$$\eqalign{
&{\cal A}_{4,1}=\cr
&\ffc(504 \nu^6+120 \nu^7-2304 \nu-2835-18 \nu^8-4 \nu^9+9216 \nu^2+
3280 \nu^3-4032 \nu^4-1092 \nu^5+2835 \eps,2903040) v_1^2+\cr
&q_1 (\ffc(\nu
 (-1+\nu) (\nu-2) (-3+\nu) (-4+\nu) (4+\nu) (3+\nu) (2+\nu) (1+\nu),362880) v_1^2)+\cr
&q_1^2 (
\ffc(-504 \nu^6+120 \nu^7-2304 \nu+2835+18 \nu^8-4 \nu^9-9216 \nu^2+3280 \nu^
3+4032 \nu^4-1092 \nu^5-2835 \eps,2903040) v_1^2)+...\cr
}$$

\noi
The predicted integrality of all coefficients in the above expansions
for any choice of $\nu \in \ZZ$ is quite impressive given the
complicated denominators. 
The expansions of the generating functions for $h=2$ are:
$$\eqalign{
{\cal A}_{0,2}=
&\ffc(-\nu (\nu+1),2) v_1 v_2+\ffc(-\eps \nu (\nu+1) (2 \nu+1),3) v_1 v_2^2
\ffc(-\nu^2 (\nu+1)^2,1) v_1^2 v_2^2+\hskip200pt\cr
&q_1 (\ffc(\nu^2,1) v_1 v_2+\ffc(\eps \nu^2 (2 \nu+1),1) 
v_1 v_2^2+\ffc(\nu^2 (2 \nu+1)^2,1) v_1^2 v_2^2)+\cr
&q_1^2 (\ffc(-\nu (-1+\nu),2) v_1 v_2
\ffc(-\eps \nu^2 (-1+2 \nu),1) v_1 v_2^2 \ffc(-6 \nu^4,1) v_1^2 v_2^2)+\cr
&q_1^3 (\ffc(0
,1) v_1 v_2+\ffc(\eps \nu (-1+2 \nu) (-1+\nu),3) v_1 v_2^2+\ffc(\nu^2 (-1+2 \nu)^2,1)
 v_1^2 v_2^2)+...\cr
}$$
$$\eqalign{
&{\cal A}_{1,2}=\cr
&\ffc(\nu (-1+\nu) (\nu+2) (\nu+1),24) v_1 v_2+\ffc(\eps \nu (\nu+1) (2 \nu+1) (\nu^2+\nu-1)
,6) v_1 v_2^2+\ffc(2 \nu^2 (2 \nu^2+2 \nu-1) (\nu+1)^2,3) v_1^2 v_2^2+\cr
&q_1 (\ffc(-\nu^2 (
-1+\nu) (\nu+1),12) v_1 v_2+\ffc(-\eps \nu^2 (2 \nu+1) (3 \nu-2) (\nu+1),6) v_1 v_2^2+\ffc
(-2 \nu^2 (\nu+1) (-1+2 \nu) (2 \nu+1)^2,3) v_1^2 v_2^2)+\cr
&q_1^2 (\ffc(\nu (-1+\nu) (\nu-2) (\nu
+1),24) v_1 v_2+\ffc(\eps \nu^2 (-1+\nu) (3 \nu+2) (-1+2 \nu),6) v_1 v_2^2+\ffc(4 \nu^4 
(-1+2 \nu^2),1) v_1^2 v_2^2)+\cr
&q_1^3 (\ffc(-\eps \nu (-1+2 \nu) (-1+\nu
) (\nu^2-\nu-1),6) v_1 v_2^2+\ffc(-2 \nu^2 (2 \nu+1) (-1+\nu) (-1+2 \nu)^2,3) v_1^2 v_2
^2)+...\cr
}$$
$$\eqalign{
&{\cal A}_{2,2}=\cr
&\ffc(-\nu (-1+\nu) (\nu-2) (\nu+3) (\nu+2) (\nu+1),720) v_1 v_2+\ffc(-\eps \nu (-1+\nu) (2
 \nu+1) (\nu+2) (\nu+1) (13 \nu^2+13 \nu-12),360) v_1 v_2^2+\cr
&q_1 (\ffc(\nu^2 (-1+\nu) (\nu-2) (\nu+2) (\nu+1)
,360) v_1 v_2+\ffc(\eps \nu^2 (-1+\nu) (2 \nu+1) (\nu+1) (39 \nu^2+26 \nu-34),360) v_1 
v_2^2)+\cr
&q_1^2
 (\ffc(-\nu (-1+\nu) (\nu-2) (\nu-3) (\nu+2) (\nu+1),720) v_1 v_2+\ffc(-\eps \nu^2 (-1+\nu) (-1+
2 \nu) (\nu+1) (39 \nu^2-26 \nu-34),360) v_1 v_2^2)+\cr
&q_1^3 (\ffc(\eps \nu (-1+\nu) (\nu-2) (-1+2 \nu) (\nu+1) (13 
\nu^2-13 \nu-12),360) v_1 v_2^2)+...
}$$
$$\eqalign{
&{\cal A}_{3,2}=\hskip400pt\cr
&\ffc(\nu (-1+\nu) (\nu-2) (\nu-3) (\nu+4) (\nu+3) (\nu+2) (\nu+1),40320) v_1 v_2+\cr
&q_1 (\ffc(-\nu^2 (-1+\nu) (\nu-2) (\nu-3) (\nu+3) (\nu+2) (\nu+1),20160) v_1 v_2)+\cr
&q_1^2 (\ffc(\nu (-1+\nu) (\nu-2) (\nu-3) (\nu-4) (\nu+3) (\nu+2) (\nu+1),40320) v_1 
v_2)+\cr
&q_1^3 (\ffc(-\eps \nu (-1+\nu) (\nu-2) (-1+2 \nu) (\nu+1)
 (205 \nu^4-410 \nu^3-614 \nu^2+819 \nu+558),45360) v_1 v_2^2)+...\cr
}$$
$$\eqalign{
&{\cal A}_{4,2}=\hskip400pt\cr
&\ffc(-\nu (-1+\nu) (\nu-2) (\nu-3) (\nu-4) (\nu+5) (\nu+4) (\nu+3) (\nu+2) (\nu+1),
3628800) v_1 v_2+\cr
&q_1 (\ffc(\nu^2 (-1+\nu) (\nu-2
) (\nu-3) (\nu-4) (\nu+4) (\nu+3) (\nu+2) (\nu+1),1814400) v_1 v_2)+\cr
&q_1^2 (\ffc(-\nu (-1+\nu) (\nu-2) (\nu-3) (\nu-4) (\nu-5) (\nu+4) (\nu+3) (\nu+2) (\nu+1)
,3628800) v_1 v_2)+...
}$$
\listrefs
\end